\documentclass[useAMS,usenatbib,onecolumn,usegraphicx]{mn2e}
\usepackage{amssymb}

\title[Spectral and algebraic instabilities in  Keplerian discs
]
{
Spectral and algebraic instabilities in thin Keplerian discs\\
under poloidal  and toroidal  magnetic fields
}
\author[Yuri M. Shtemler, Michael Mond, and Edward Liverts]
{Yuri M. Shtemler$^{1}$\thanks{E-mail:
shtemler@bgu.ac.il; mond@bgu.ac.il; eliverts@bgu.ac.il},  Michael Mond$^{1}$, and  Edward  Liverts$^{1}$\\
$^{1}$Department of Mechanical Engineering,  Ben-Gurion
University of the Negev,  P.O. Box 653, Beer-Sheva 84105,
Israel
}
\begin{document}

\date{Accepted ---. Received ----; in original form ----}

\pagerange{\pageref{firstpage}--\pageref{lastpage}} \pubyear{}

\maketitle

\label{firstpage}

\begin{abstract}
Linear instability of two equilibrium configurations with either poloidal (I) or toroidal (II) dominant magnetic field components are studied in thin vertically-isothermal Keplerian discs. Solutions of the stability problem are found explicitly by asymptotic expansions in the small aspect ratio of the disc. In both equilibrium configurations the perturbations are decoupled into in-plane and vertical modes.  For equilibria of type I those two modes are the Alfv\'en-Coriolis and sound waves, while for equilibria of type II they are the inertia-Coriolis and magnetosonic waves. Exact expressions for the growth rates as well as the number of unstable modes for type I equilibria are derived. Those are the discrete counterpart of the continuous infinite homogeneous cylinder magnetorotational (MRI) spectrum. It is further shown that the axisymmetric MRI is completely suppressed by dominant toroidal magnetic fields (i.e. equilibria of type II). This renders the system prone to either non-axisymmetric MRI or non-modal algebraic growth mechanisms. The algebraic growth mechanism investigated in the present study occurs exclusively due to the rotation shear, generates the inertia-Coriolis driven magnetosonic modes due to non-resonant or resonant coupling that induces, respectively, linear or quadratic temporal growth of the perturbations.
\end{abstract}

\section{Introduction}
The stability properties of Keplerian disks have been a focus of intensive investigation of theoretical astrophysicists over the last decades, pertaining to the problem of angular momentum transfer in accretion disks.  Traditionally, two routes have been taken by researchers in order to elucidate various aspects of that topic.
Thus, the results of the analytical study of magneto-rotational instability (MRI) in an infinitely long cylinder [\cite{Velichov 1959};  \cite{Chandrasechar 1960}] have been  adopted for the thin disks   in order to derive criteria for the spectral stability under various conditions [\cite{Balbus and Hawley 1991}]. On the second route, numerical calculations have been performed, usually by employing the shearing-box model, in order to study the nonlinear evolution of small perturbations, and their subsequent development into full sustainable turbulence [\cite{Brandenburg et al. 1995}; \cite{Hawley et al. 1996}; \cite{Bodo et al. 2008}; \cite{Regev and Umurhan 2008}]. The main thrust of both routes has been to establish the  MRI as the main generator of sustainable turbulence. Notwithstanding accumulating analytical as well as numerical results, the notion of the MRI as a driver of turbulence and angular momentum transport has recently been shown to suffer from non-trivial difficulties in the sheer use of the shearing box in the simulations [\cite{Bodo et al. 2008}; \cite{Regev and Umurhan 2008}] as well as in doubts regarding the numerical convergence and resolution [\cite{Lesur and Longaretti 2007}; \cite{Fromang and Papaloizou 2007}; \cite{Fromang et al. 2007};  \cite{Pessah et al. 2007}]. In addition, it has recently been shown that if the thin disk geometry is taken into account both the growth rates as well as the number of unstable MRI modes are greatly reduced and are decreasing functions of the disk thickness [ \cite{Coppi and Keyes  2003};  \cite{Liverts and Mond  2009}]. In parallel, an alternative route that has recently emerged within which the dynamical response of the thin disks is investigated asymptotically in the small thickness-to-radius ratio has been shown to lead to fruitful and physically sound results. In particular, applying such strategy it has been shown that algebraic non-modal instabilities are the primary source of pure hydrodynamical activity [\cite{Umurhan et al. 2006}; \cite{Rebusco et al. 2009}; \cite{Shtemler et al. 2010}]. Furthermore, the mechanism for such non-exponential growth has been identified as the resonant as well as non-resonant interaction between inertia-Coriolis modes and sound waves [\cite{Shtemler et al. 2010}]. Interestingly enough, such an investigation has never been carried out for magnetized disks that are described by the magnetohydrodynamic (MHD) model. It is thus one of the main goals of the current work to carry out a comprehensive study of the MHD response of thin Keplerian disks taking into account their true thin-disk geometry.

In that connection note that although the magnetic field configuration in real discs is largely unknown, in the initial stage of the disc rotation the poloidal field is commonly accepted as the dominant component. However, observations and numerical simulations indicate that toroidal magnetic field may be of the same order of magnitude, or even dominate the poloidal magnetic field [see e.g. \cite{Terquem and Papaloizou 1996}; \cite{Papaloizou and Terquem  1997}; \cite{Hawley and Krolik 2002}; \cite{Proga  2003}]. In particular, the unstable modes give rise to MRI [\cite{Balbus and Hawley 1991}] which amplifies considerably the toroidal component of the magnetic field to the point that it may dominate the original axial field [\cite{Pessah and Psaltis 2005}; \cite{Begelman and Pringle 2007}]. It is of great importance therefore to investigate how the magnetic field topology influences the stability properties of the disc. The way to do that is to follow the linear problem in terms of a toroidal dominated magnetic field instead of poloidal one leading to MRI. Thus, one more goal of the present work is investigation of two types of magnetic field equilibrium configurations with either poloidal or toroidal dominated magnetic field component and their relations with spectral and non-modal mechanisms of instability in thin discs.

The paper is organized as follows. The dimensionless governing equations and their approximation to leading order in small aspect ratio of the steady-state disc are presented in the next Section. Two magnetic field configurations with dominated axial (I) and dominated toroidal (II) equilibrium magnetic fields are also described in that Section. Sections 3 and 4 describe the principal hydro-magnetic modes which can propagate in thin Keplerian discs for I and II types of equilibria, respectively. Summary and discussion are presented in Section 5. In Appendix A the general relations for comparable in-plane components of the equilibrium magnetic field are discussed in the limit of large plasma beta, and the qualitative stability analysis is made.

\section{THE PHYSICAL MODEL FOR THIN KEPLERIAN DISCS}
The stability of radially as well as axially stratified rotating plasma in thin vertically isothermal discs threaded by a magnetic field is considered. Viscosity, electrical resistivity, and radiation effects are ignored.
\subsection{Governing equations}
As a first step, all physical variables are transformed to non-dimensional quantities by using the following characteristic values  [\cite{Shtemler et al. 2009}; \cite{Shtemler et al. 2010}]:
$$
t_*=\frac{1}{\Omega_*},\,\,V_*=\frac{r_*}{t_*},\,\,
L_*=V_* t_*,\,\,
 m_*= m_i,\
n_*=n_i,\,\,\,
\,\,\,\,\,\,\,\,\,\,\,\,\,\,\,\,\,\,\,\,\,\,\,\,\,\,\,\,\,\,\,\,\,\,\,\,\,\,\,\,\,\,\,\,\,\,\,\,\,\,\,\,\,\,\,\,\,\,\,\,
\,\,\,\,\,\,\,\,\,\,\,\,\,\,\,\,\,\,\,\,\,\,\,\,\,\,\,\,\,\,\,\,\,\,\,\,\,\,\,\,\,\,\,\,\,\,\,\,\,\,\,\,\,\,\,\,\,\,\,\,
\,\,\,\,\,\,\,\,\,\,\,\,\,\,\,\,\,\,\,\,\,\,\,\,\,\,\,\,\,\,\,\,\,\,\,\,\,\,\,\,\,\,\,\,\,\,\,\,\,\,\,\,\,\,\,\,\,\,\,\,
$$
\begin{equation}
\Phi_*={V_*}^2,\,\,
c_{S*}=\sqrt{T_*/m_*},\,\,\,P_*=m_* n_* c_{S*}^{\,\,2},\,\,
j_*=\frac{c}{4\pi}\frac{B_*}{r_*},\,\,
E_*=\frac{V_* B_*}{c}.\,\,
\,\,\,\,\,\,\,\,\,\,\,\,\,\,\,\,\,\,\,\,\,\,\,\,\,\,\,\,\,\,
\label{1}
\end{equation}
Here $\Omega_*=(GM_c/r_*^3)^{1/2}$  is the Keplerian angular
velocity of the fluid at the characteristic radius $r_*$ that
belongs to the Keplerian portion of the disc;
$G$  is the gravitational constant; $M_c$ is the total mass of the
central object; $c$ is the speed of light; $\Phi_*$  is the characteristic value of the
gravitational potential; the characteristic
mass and number density equal to
%
 %
 the ion mass and number density, $m_*=m_i$  and $n_*=n_i$.
%
The characteristic values of the electric current density and electric field, $j_*$
and $E_*$   have been chosen consistently with Maxwell's equations; $c_{S*}$ is the characteristic
sound velocity; $T_*=T(r_*)$  is the characteristic temperature;
the characteristic magnetic field is specified below depending on the magnetic field configuration:
 $B_*=B_z(r_*)$
 if the axial equilibrium magnetic field dominates or of the order toroidal one, and  $B_*=B_\theta(r_*)$  otherwise;
the dimensional equilibrium temperature and the two components of the equilibrium  magnetic field, $T(r)$  and  $B_z(r)$,   $B_\theta(r)$, respectively, are free functions.

The resulting dimensionless dynamical equations for vertically isothermal discs are:
\begin{equation}
\frac{ D \bf {V} }{Dt} = -\frac{ 1}{M_S^2}\frac{\nabla P}{n}-  \nabla \Phi+\frac{1}{\beta M^2_S} \frac{\bf{j}
\times {\bf{B}}}{n},\\
\label{2}
\end{equation}
\begin{equation}
\frac{\partial n} {\partial t}
               + \nabla \cdot(n {\bf{V}} ) =0,\\
\label{3}
\end{equation}
\begin{equation}
\frac{\partial {\bf{B} }} {\partial t}+
 \nabla \times
 {\bf{E}}=0,\,\nabla \cdot {\bf{B}}=0,
\label{4}
\end{equation}
\begin{equation}
 {\bf {E} }=  -  {\bf{V}} \times {\bf{B}},
\label{5}
\end{equation}
\begin{equation}
P =nT.
\label{6}
\end{equation}
Here $\nabla P=\bar{c}^2_S\nabla n$  for vertically isothermal discs, and the dimensionless equilibrium sound
speed is given by $\bar{c}^2_S =\partial P/\partial n\equiv T(r)$. Standard cylindrical coordinates
$\{r,\theta,z\}$   are adopted throughout the paper with the associated unit vectors
$\{\bf{i}_r,\bf{i}_\theta,\bf{i}_z\}$; $\bf{V}$  is the plasma velocity; $t$ is time; $D/Dt=\partial/\partial
t+(\bf{V}\cdot\nabla)$  is the material derivative; $\Phi(r,z)=-(r^2+z^2)^{-1/2}$  is the
 gravitational potential due to the central object; $\bf{B}$
   is the magnetic field, $\bf{j}=\nabla\times\bf{B}$  is the current density; $\bf{E}$
  is the electric field; $P=P_e+P_i$  is the total plasma pressure; $P_l=n_lT_l$  are the partial species
  pressures ($l=e,i$); $T=T_e=T_i$  is the plasma temperature; subscripts $e$ and $i$ denote electrons
  and ions, respectively. Note that a preferred direction is tacitly defined here, namely, the
  positive direction of the $z$ axis is chosen according to positive Keplerian rotation. The dimensionless
  coefficients $M_S$  and $\beta$  are the Mach number and the characteristic plasma beta, respectively:
\begin{equation}
M_S=\frac{V_*} {c_{S*}},\,\, \beta=\frac{P_*} {B_{*}^2}.
\label{7}
\end{equation}
Zero conditions at infinity for the in-plane magnetic field are adopted, namely:
\begin{equation}
B_r=0,\,\,\, B_\theta=0\,\,\,\, \,\,\,\, \mbox{for} \,\,\,z=\pm \infty.
\label{8}
\end{equation}
The boundary condition for the axial velocity will be specified below as the least possible divergence at infinity. As will be seen later on, despite such unbounded growth the axial mass flux tends to zero far away from the mid-plane due to the fast vanishing of number density at  $z=\pm \infty$. To simplify the further treatment of Maxwell's equations, both the hydro-magnetic basic configuration and perturbations are assumed to be axisymmetric.

A common property of thin Keplerian discs is their highly
compressible motion with large Mach numbers [ \cite{Frank et al.
2002}]. Furthermore, the characteristic effective semi-thickness of the equilibrium disc $H_* = H(r_*)\,\, (H =
H(r)$ is the local semi-thickness) is defined so that the disc aspect ratio $\epsilon$  equals the inverse
Mach number:
\begin{equation}
\frac{1}{M_S}=\epsilon=\frac{H_*}{r_*}
\ll 1.
\label{9}
\end{equation}
Thus, the thin disc approximation means
\begin{equation}
\frac{1}{M_S}=\sqrt{\frac{r_*T_*}{GM_c}}=\epsilon\ll 1,\,\,
\label{10}
\end{equation}
where
$$
M_S=\frac{V_*}{c_{S*}},\,\,\,V_*=r_*\Omega_*,\,\,\*\Omega_*=\sqrt{\frac{GM_c}{r_*^3}},
\,\,\, c_{S*}=\sqrt{T_*},\,\,\, H_*= \frac{c_{S*}}{\Omega_*} \,.
\,\,\,\,\,\,\,\,\,\,\,\,\,\,\,\,\,\,\,\,\,\,\,\,\,\,\,\,\,\,\,\,\,\,\,\,\,\,\,\,\,\,\,\,\,\,\,\,\,\,
\,\,\,\,\,\,\,\,\,\,\,\,\,\,\,\,\,\,\,\,\,\,\,\,\,\,\,\,\,\,\,\,\,\,\,\,\,\,\,\,\,\,\,\,\,\,\,\,\,\,
\,\,\,\,\,\,\,\,\,\,\,\,\,\,\,\,\,\,\,\,\,\,\,\,\,\,\,\,\,\,\,\,\,\,\,\,\,\,\,\,\,\,\,\,\,\,\,\,\,\,
$$
The smallness of $\epsilon$  means that dimensionless axial coordinate is also small, i.e. $z/r_*\sim
\epsilon\,\,({\mid} z{\mid}  ^{<}_\sim H_*)$, and consequently the following rescaled quantities may be introduced in
order to
further apply the asymptotic expansions in $\epsilon $  [similar to Shtemler et al. (2009); Shtemler et al. (2010)]:
\begin{equation}
\zeta=\frac{z}{\epsilon}\sim\epsilon^0,\,\,\,\bar{H}(r) =\frac{H(r)}{\epsilon}\sim\epsilon^0,
\label{11}
\end{equation}
where $\bar{H}(r)=\bar{c}_S(r)/\bar{\Omega}(r)$  is the scaled semi-thickness of the disc.

\subsection{Equilibrium configurations}
We start by deriving the steady state solution. It is first noted that the asymptotic expansion
for the time-independent gravitational potential is given by:
\begin{equation}
\Phi(r,\zeta)=\bar{\Phi}(r)+\epsilon^2\bar{\phi}(r,\zeta),\,\,\,
\bar{\Phi} (r)=-\frac{1}{r},\,\,\, \bar{\phi}(r,\zeta)
=\frac{1}{2}\zeta^2\bar{\Omega}^2(r)+O(\epsilon^2),\,\,\, r>1>>\epsilon.
\label{12}
\end{equation}
Substituting (\ref{12}) into Eqs. (\ref{2})-(\ref{6}) and  setting the partial derivatives with respect to time to zero yield
to leading order in $\epsilon$
 \begin{equation}
\frac{\bar{V}_\theta^2}{r}=\frac{d\bar{\Phi}(r)}{dr},\,\,\,\,
\frac{\bar{c}_S^2(r)}{\bar{n}}\frac{\partial \bar{n}}{\partial \zeta}=
-\frac{\partial \bar{\phi}(r,\zeta)}{ \partial
\zeta}.
\label{13}
\end{equation}
Thus, the  velocity as well as the number density are as follows:
\begin{equation}
V_r=o(\epsilon^2),\,\,\, V_\theta =\epsilon ^0\bar{V}_\theta(r)+O(\epsilon ^2),\,\,\,
V_z=o(\epsilon ^2),\,\,\, n\cong\epsilon ^0 \bar{n}\equiv \epsilon ^0\bar{N}(r)\bar{\nu}(\eta),
\label{14}
\end{equation}
where $o(\epsilon)\ll \epsilon, \, O(\epsilon)\sim \epsilon$, and
\begin{equation}
\bar{V}_\theta(r)=r\bar{\Omega} (r),\,\,\, \bar{\Omega} (r)=r^{-3/2},\,\,\,
\bar{\nu}( \eta )=\exp(-\eta ^2/2), \ \ \ \eta=\zeta/\bar{H}(r).
\label{15}
\end{equation}

Two quite different equilibrium magnetic configurations will be considered below, namely one with comparable magnitudes of the axial and toroidal components of the magnetic field
(mainly studied for the dominant poloidal magnetic field, while the general case of comparable components of the magnetic field is discussed in Appendix A),
 and the other with dominant toroidal component. The two equilibria (denoted I and II below) are distinguished by different scaling of the physical variables with   $\epsilon$ which generally may be written as
$f(r,\zeta)= \epsilon^{\bar{S}} \bar{f}(r,\zeta)$. All equilibrium variables are written in the leading order in  $\epsilon$, and depend on the radial variable only. The exceptions are the number density and the pressure that depend on the axial coordinate in a self-similar manner with radius-dependent amplitude. The toroidal and axial magnetic fields as well as the disc thickness and the amplitude factor, $\bar{N}(r)$, in the number density are arbitrary functions of the radial variable. Those functions specify the equilibrium state.

To start the equilibrium description it is first assumed that both types of equilibria under the current study are characterized by a toroidal component of the magnetic field that is of order  $\epsilon^0$.
In contrast, in one equilibrium system (I) the axial component of the magnetic field is of order   $\epsilon^0$  as well, while in the other  system (II) it is of order
$\epsilon$. These assumptions together with relations (\ref{12})-(\ref{15}), determine the order in $\epsilon$  of the rest of the physical variables:

{\it{(I) Comparable equilibrium magnetic field components}}
\begin{equation}
B_r = o(\epsilon^2),\,\,\, B_\theta \cong\epsilon^0 \bar{B}_\theta(r),\,\,\, B_z \cong\epsilon ^0 \bar{B}_z(r),
j_r = o(\epsilon^2),\,\,\,
j_ \theta \cong\epsilon ^0\bar{j}_ \theta(r)\equiv -\epsilon ^0 \frac{d \bar{B}_z}{d r},\,\,\,
j_z \cong\epsilon ^0 \bar{j}_z \equiv \epsilon ^0 \frac{1}{r} \frac{d (r\bar{B}_\theta)}{d r}.\,\,\,
\label{16}
\end{equation}

 {\it{(II) Dominant equilibrium toroidal magnetic field }}
 \begin{equation}
 B_r = o(\epsilon^2),\,\,\, B_\theta \cong\epsilon^0 \bar{B}_\theta(r),\,\,\, B_z \cong\epsilon  \bar{B}_z(r),
j_r = o(\epsilon^2),\,\,\,
j_\theta \cong\epsilon \bar{j}_\theta(r)\equiv -\epsilon  \frac{d \bar{B}_z}{d r},\,\,\,
j_z \cong\epsilon ^0 \bar{j}_z \equiv \epsilon ^0 \frac{1}{r} \frac{d (r\bar{B}_\theta)}{d r}.\,\,\,
\label{17}
\end{equation}

For convenience the results are summarized in Table 1.
It is noted finally that to lowest order in $\epsilon$  the magnetic field configurations under consideration do not influence the steady-state properties of the disk. As will be seen in the following sections, this situation changes dramatically when small perturbations are considered.

 \begin{table*}
 \centering
 \begin{minipage}{140mm}
\caption{Scales in $\epsilon$  of the equilibrium variables for two types of equilibria (I) and  (II).
}
\begin{tabular}{@{}ccccccccccc@{}}
$f=\epsilon^{\bar{S}}\bar{f}$
& $n$ & $V_r$ & $V_\theta$ & $V_z$
& $B_r$ & $B_\theta$ & $B_z$ & $j_r$ &  $j_\theta$ & $j_z$
    \\
      \hline
       I, $\bar{S}$
       & $0$  &  $>2$  &   0  & $>2$
              &  $>2$  &   0  &  0
              &  $>2$  &   0  & 0
              \\
        \hline
       II, $\bar{S}$
       & $0$ & $>2$    & 0     & $>2$
             & $>2$    & 0     & 1
             & $>2$    & 1     & 0
            \\
  \end{tabular}
\end{minipage}
\end{table*}

%

\subsection{Perturbed thin discs}
In general for the unsteady nonlinear case the dependent variables are scaled in $\epsilon$  in the following way:
\begin{equation}
f(r,\zeta,t)=\epsilon^{\bar{S}}\bar{f}(r,\zeta)+\epsilon^{S'} f '(r,\zeta,t).
\label{18}
\end{equation}
Here $f(r,\zeta,t)$  stands for any dependent variable, the bar and the prime denote equilibrium and
 perturbed variables; each perturbed variable is characterized by some power
 $S'$ that is different for two types of equilibria I and II.
The various cases are summarized in (Table 2).

\section{DYNAMICAL EQUATIONS FOR THE PERTURBED DISC: TYPE I EQUILIBRIA}
Consider the general dynamic equations for the perturbed disc equilibria of type I, namely thin discs under the effect of a magnetic field with comparable poloidal and toroidal components. Then the detailed stability study is carried out for zero toroidal magnetic field, while a qualitative analysis of the comparable poloidal and toroidal components is presented in Appendix A.

\subsection{ The reduced nonlinear equations}
Substituting (\ref{12})-(\ref{18}) into (\ref{2}) - (\ref{6}) yields to leading order in $\epsilon$ the following system of the reduced
 non-linear equations that governs the dynamical behavior of thin discs:
\begin{equation}
\frac{\partial V_r'}{\partial t}
-2\bar{\Omega}(r)V_\theta'
+V_z'\frac{\partial V_r'}{\partial \zeta}
-\frac{1}{\beta}\frac{\bar{B}_z(r)}{\bar{n}+n'}\frac{\partial B_r'}{\partial \zeta }  =0,
\label{19}
\end{equation}
\begin{equation}
\frac{\partial V_\theta' }{\partial t}
+\frac{1}{r}\frac{d (r^2 \bar{\Omega})}{d r}V_r'
+V_z'\frac{\partial V_\theta'}{\partial \zeta}
-\frac{1}{\beta}\frac{\bar{B}_z(r)}{\bar{n}+n'}\frac{\partial B_\theta'}{\partial \zeta } =0,
\label{20}
\end{equation}
\begin{equation}
\frac{\partial  V_z'}{\partial t}
+\bar{c}_S^2(r)\frac{\bar{n}}{\bar{n}+n'}
\frac{\partial}{\partial \zeta }(\frac{n'}{\bar{n}})
+\frac{1}{\beta}\bar{B}_\theta(r)\frac{\partial}{\partial \zeta }(\frac{B_\theta'}{\bar{n}+n'})=
-\frac{1}{2}\frac{\partial}{\partial \zeta}
(V_z^{'2}+\frac{1}{\beta}\frac{B_\theta^{'2}+B_r^{'2}}{\bar{n}+n'}),
\label{21}
\end{equation}
\begin{equation}
\frac{\partial n'}{\partial t}
+\frac{\partial [(\bar{n}+n')V_z']}{\partial \zeta }=0,
\label{22}
\end{equation}
\begin{equation}
\frac{\partial B_r'}{\partial t}
-\bar{B}_z(r)\frac{\partial V_r'}{\partial \zeta}=
-\frac{\partial V_z' B_r'}{\partial \zeta },
\label{23}
\end{equation}
\begin{equation}
\frac{\partial B_\theta'}{\partial t}
-\bar{B}_z(r)\frac{\partial V_\theta'}{\partial \zeta}
-r\frac{d \bar{\Omega}}{d r}B_r'=
-\frac{\partial V_z' B_\theta'}{\partial \zeta }
-\bar{B}_\theta(r)\frac{\partial V_z' }{\partial \zeta }.
\label{24}
\end{equation}
The system of equations (\ref{19}) - (\ref{24}) is subject to the boundary condition of least possible divergence for the axial velocity at infinity along with the conditions for the in-plane magnetic field as follows  from
 (8)
\begin{equation}
B_r'=0,\,\,\,B_\theta'=0\,\,\,\,\,\,\,\, \mbox{for} \,\,\,\,\,\,\zeta=\pm \infty.
\label{25}
\end{equation}
%
Both poloidal and toroidal components of the perturbed magnetic field, $B_r'$  and   $B_z'$ , are expressed through the magnetic flux function,   $\Psi'$, which provides the divergent free magnetic field,  $\nabla \cdot {\bf{B}}=0$. Here the equation for  $B_r'$  is obtained by differentiating by  $\zeta$  the equation for  $\Psi'$, while the corresponding
 equation for the perturbed axial magnetic field, $B'_z$,   decouples from the rest equations, and drops out
from the governing system of equations (\ref{19}) - (\ref{24}). Equation (\ref{21}) has been derived with
the help of the second steady-state relation (\ref{14}). Furthermore, the radial derivatives drop
out from the system of Eqs. (\ref{19}) - (\ref{25}). This, it should be stressed,  is so not due to a frozen
coefficients assumption, but is a direct result of the thin disc geometry. An important conclusion from the
absence of radial derivatives in the  reduced system of equations of the thin disc approximation is that transient non-exponential growth
is not possible in such configuration. As will subsequently be
shown (see  Section 4 of the present paper), this picture changes dramatically when the dominant magnetic fields are toroidal, in which case the only possible axisymmetric instability is due to non-modal algebraic transient growth.

 \begin{table*}
 \centering
 \begin{minipage}{140mm}
\caption{Scales in $\epsilon$  of the perturbed variables for two types of equilibria (I) and  (II). }
\begin{tabular}{@{}ccccccccccc@{}}
$f=\epsilon^{\bar{S}}\bar{f}+\epsilon^{S'} f'$
& $n$ & $V_r$ & $V_\theta$ & $V_z$
& $B_r$ & $B_\theta$ & $B_z$
& $j_r$ &  $j_\theta$ & $j_z$
    \\
        \hline
       I, $S'$
       & $0$ & $1$   &   1     & $1$
             & $0$   &   0     & 1
             & $-1$  &  $-1$   & 0
        \\
         \hline
       II, $S'$
       & $0$ & $0$   & 0    & $1$
             & $0$   & 0    & 1
             & $-1$  & $-1$
             & 0
         \\
\end{tabular}
\end{minipage}
\end{table*}

As the radial coordinate is a mere parameter in the set of the reduced equations, it is convenient to replace the physical variables by the following self-similar quantities:
\begin{equation}
\tau=t,\,\,\,\ \,\,\,\eta=\frac{\zeta}{\bar{H}(r)},\,\,\,
\label{26}
\end{equation}
such that the derivatives in the new and old variables are related as follows:
\begin{equation}
\frac{\partial }{\partial t} = \frac{\partial }{\partial \tau },\,\,\,
\frac{\partial }{\partial \zeta} =\frac{1}{\bar{H}(r)} \frac{\partial }{\partial \eta }.\,\,\,
\label{27}
\end{equation}
%
Finally, system (\ref{19}) - (\ref{25}) may be written in a simpler form by introducing the following scaled
dependent variables:
\begin{equation}
{\bf{v}}(\tau,r,\eta)=\frac{ { \bf{V}  }'}{\bar{c}_S (r)},\,\,\,
\nu(\tau,r,\eta)=\frac{ { n'  }}{\bar{N}(r)},\,\,\,
{\bf{b}}(\tau,r,\eta)=\frac{ { \bf{B}  }'}{\bar{B}_z (r)}. \,\,\,
\label{28}
\end{equation}
Below for convenience and with no confusion the  notation  $t$ for the time variable is reinstated instead of the new variable $\tau$.
This yields the following system of equations that depend parametrically on the radius:
\begin{equation}
\frac{1}{\bar{\Omega}(r)} \frac{\partial v_r}{\partial t}
-2v_\theta
-\frac{1}{\bar{\beta}_z(r)}\frac{1}{\bar{\nu}(\eta)+\nu}\frac{\partial b_r}{\partial \eta }  =
-v_z\frac{\partial v_r}{\partial \eta},
\label{29}
\end{equation}
\begin{equation}
\frac{1}{\bar{\Omega}(r)}\frac{\partial v_\theta }{\partial t}
+\frac{1}{2}v_r
-\frac{1}{\bar{\beta}_z(r)}\frac{1}{\bar{\nu}(\eta)+\nu}\frac{\partial b_\theta}{\partial \eta }
=-v_z\frac{\partial v_\theta}{\partial \eta},
\label{30}
\end{equation}
\begin{equation}
\frac{1}{\bar{\Omega}(r)}\frac{\partial  v_z}{\partial t}
+\frac{\bar{\nu}(\eta)}{\bar{\nu}(\eta)+\nu}
\frac{\partial}{\partial \eta }(\frac{\nu}{\bar{\nu}(\eta)})
+\frac{\bar{S}(r)}{\bar{\beta}_z(r)}
\frac{\partial}{\partial \eta }[\frac{b_\theta}{\bar{\nu}(\eta)+\nu}]=
-\frac{1}{2}\frac{\partial}{\partial \eta}
[v_z^2+\frac{1}{\bar{\beta}_z(r)}\frac{b_\theta^2+b_r^2}{\bar{\nu}(\eta)+\nu}],
\label{31}
\end{equation}
\begin{equation}
\frac{1}{\bar{\Omega}(r)}\frac{\partial \nu}{\partial t}
+\frac{\partial [(\bar{\nu}(\eta)+\nu]v_z}{\partial \eta }=0,
\label{32}
\end{equation}
\begin{equation}
\frac{1}{\bar{\Omega}(r)}\frac{\partial b_r}{\partial t}
-\frac{\partial v_r}{\partial \eta}=
-\frac{\partial (v_z b_r)}{\partial \eta },
\label{33}
\end{equation}
\begin{equation}
\frac{1}{\bar{\Omega}(r)}\frac{\partial b_\theta}{\partial t}
-\frac{\partial v_\theta}{\partial \eta}
+\frac{3}{2} b_r=
-\frac{\partial (v_z b_\theta)}{\partial \eta }
-\bar{S}(r)
\frac{\partial v_z}{\partial \eta },
\label{34}
\end{equation}
supplemented by the boundary condition  of  least possible divergence  for the axial velocity along with the  vanishing conditions for the in-plane magnetic field at infinity.
Here the Keplerian epicyclical frequency,
$\bar{\chi}(r)=\bar{\Omega} (r)$ has been employed; $\bar{\nu} (\eta)$  is the scaled equilibrium density;
$\bar{\beta}_z (r)$  and $\bar{\beta}_\theta (r)$  are the local axial and toroidal plasma beta functions:
\begin{equation}
\bar{\beta}_z(r)=\beta\frac{\bar{N} (r) \bar{c}_S^2 (r) }{\bar{B}_z^2 (r) },\,\,\,\,\,\,\,
\bar{\beta}_\theta (r) =\beta\frac{\bar{N}(r) \bar{c}_S^2 (r) }{\bar{B}_\theta^2 (r) },\,\,\,\,\,\,\,
\bar{S(r)}=\sqrt{\frac{\bar{\beta}_z(r)}{\bar{\beta}_\theta(r)}}\equiv\frac{\bar{B}_\theta(r)}{\bar{B}_z(r)},\,\,\,
\label{36}
\end{equation}
where both local parameters  $\bar{\beta}_z(r)$ and  $\bar{\beta}_\theta(r)$  are proportional to the characteristic plasma  beta, $\beta $.

Relations (\ref{29})-(\ref{36}) form the full nonlinear MHD problem in the thin disc approximation and
 are named as defined above, the reduced nonlinear equations.


\subsection{The linear problem.}
Assuming now that the perturbations are small, the system of equations  (\ref{29})-(\ref{34}) may be linearized about the steady-state equilibrium solution. This yields:
\begin{equation}
\frac{1}{\bar{\Omega}(r)}\frac{\partial v_r}{\partial t}
-2v_\theta
-\frac{1}{\bar{\beta}_z(r)\bar{\nu}(\eta)}\frac{\partial b_r}{\partial \eta }  =
0,
\label{37}
\end{equation}
\begin{equation}
\frac{1}{\bar{\Omega}(r)}\frac{\partial v_\theta }{\partial t}
+\frac{1}{2}v_r
-\frac{1}{\bar{\beta}_z(r)\bar{\nu}(\eta)}\frac{\partial b_\theta}{\partial \eta }
=0,
\label{38}
\end{equation}
\begin{equation}
\frac{1}{\bar{\Omega}(r)}\frac{\partial b_r}{\partial t}
-\frac{\partial v_r}{\partial \eta}=0,
\label{39}
\end{equation}
\begin{equation}
\frac{1}{\bar{\Omega}(r)}\frac{\partial b_\theta}{\partial t}
-\frac{\partial v_\theta}{\partial \eta}
+\frac{3}{2} b_r=
-
\bar{S}(r)
\frac{\partial v_z}{\partial \eta },
\label{40}
\end{equation}
\begin{equation}
\frac{1}{\bar{\Omega}(r)}\frac{\partial  v_z}{\partial t}
+
\frac{\partial}{\partial \eta }(\frac{\nu}{\bar{\nu}(\eta)})=
-\frac{\bar{S}(r)}{\bar{\beta}_z(r)}
\frac{\partial}{\partial \eta }[\frac{b_\theta}{\bar{\nu}(\eta)}],
\label{41}
\end{equation}
\begin{equation}
\frac{1}{\bar{\Omega}(r)}\frac{\partial \nu}{\partial t}
+\frac{\partial [\bar{\nu}(\eta)v_z]}{\partial \eta }=0.
\label{42}
\end{equation}
For definiteness the characteristic magnetic field is specified as the dimensional axial component of the equilibrium magnetic field  $B_*=B_z(r_*)$.

The general case of comparable poloidal and toroidal components of the equilibrium magnetic field is presented in Appendix A for perturbations of type I. A qualitative analysis is done of the influence of the  equilibrium toroidal magnetic field on the disc stability in the limit of large plasma beta. In the case of small plasma beta ($\beta \lesssim 1$) two principal modes, namely the Alfv\'en-Coriolis (AC) and the magnetosonic (MS) modes, are strongly coupled, while in the limit of large plasma beta, AC mode decouples on its characteristic scale from the MS mode. This leaves the resulting dispersion relation the same as in the case of zero toroidal magnetic field, and the influence of the toroidal magnetic field is reduced to excitation of the AC-driven MS mode. On the characteristic scale of MS's mode, a stable MS mode decouples from the AC mode and the influence of the  equilibrium toroidal magnetic field is reduced to excitation of the stable MS-driven AC mode.

To simplify the  analysis,  the equilibrium toroidal magnetic field is set  to zero for type I equilibria, or equivalently,   $\bar{\beta}_\theta (r)=\infty$. Under such conditions the reduced linearized system  (\ref{37}) - (\ref{42})
is divided into two decoupled sub-systems that describe the dynamics of two different modes, namely: the Alfv\'en-Coriolis and the sound modes.
It is indeed the approximation that is adopted in the remaining of the current analysis of  type I equilibria.


\subsection{Linear stability analysis for the Alfv\'en-Coriolis modes.}
We start by representing the perturbations up to a radius-dependent amplitude factor as follows:
\begin{equation}
f(r,\eta,t)=\exp[-i \lambda(r)t ] \hat{f}(r,\eta),\,\,\,
\label{43}
\end{equation}
where $\lambda(r)$ is the complex eigenvalue
\begin{equation}
\lambda=  \Lambda+i\Gamma.
\label{44}
\end{equation}
%
%

Substituting (\ref{43})-(\ref{44}) into (\ref{37})-(\ref{42}) results in the following system of linear ordinary differential equations for the perturbed velocity as well as in-plane magnetic field components.
That system of equations    characterizes the Alfv\'en-Coriolis waves and depends parametrically  on the radius:
\begin{equation}
-i\lambda \hat{v}_r
-2\hat{v}_\theta
-\frac{1}{\bar{\beta}_z(r)\bar{\nu}(\eta)}\frac{d \hat{b}_r}{d \eta }  =
0,
\label{45}
\end{equation}
\begin{equation}
-i\lambda \hat{v}_\theta
+\frac{1}{2}\hat{v}_r
-\frac{1}{\bar{\beta}_z(r)\bar{\nu}(\eta)}\frac{d \hat{b}_\theta}{d \eta }
=0,
\label{46}
\end{equation}
\begin{equation}
-i\lambda \hat{b}_r
-\frac{d \hat{v}_r}{d \eta}=0,
\label{47}
\end{equation}
\begin{equation}
-i\lambda \hat{b}_\theta
-\frac{d \hat{v}_\theta}{d \eta}
+\frac{3}{2} \hat{b}_r=
0.
\label{48}
\end{equation}
In addition, the Alfv\'en-Coriolis sub-system (\ref{45})-(\ref{48}) is subject to the vanishing boundary conditions for the in-plane magnetic-field components.

The linear set of equations (\ref{45})-(\ref{48}) may be reduced to the following single fourth order
ordinary differential equations for both  $\hat{b}_r$ and $\hat{b}_\theta$:
\begin{equation}
\frac{d}{d\eta} \left[\frac{1}{\bar{\nu}(\eta)}
\frac{d^2}{d\eta^2} \left(\frac{1}{\bar{\nu}(\eta)}\frac{d\hat{b}_{r,\theta}}{d\eta}\right)\right]
+(3+2\lambda^2)\bar{\beta}_z(r)
\frac{d}{d\eta}  \left(\frac{1}{\bar{\nu}(\eta)}\frac{d\hat{b}_{r,\theta}}{d\eta}\right)
+ \lambda^2( \lambda^2-1)\bar{\beta}_z^2(r) \hat{b}_{r,\theta}=0.
\label{50}
\end{equation}
Equation (\ref{50}) is the same as the one used by \cite{Liverts and Mond 2009}   who have derived it for a model
problem under the assumption of zero radial variations of the perturbations. Here, it is important to emphasize
however that the radial coordinate is a parameter a fact that renders the radial
dependence of the perturbations arbitrary.
\cite{Liverts and Mond 2009} have solved (\ref{50}) with the aid of the Wentzel-Kramers-Brillouin (WKB) approximation for $\bar{\nu}(\eta)=\exp(-\eta^2/2)$.  Remarkably, however,   that a full analytical solution of Eq. (\ref{50}) is possible for a slightly modified density profile. The first and main step towards that goal is to replace the isothermal density vertical steady-state distribution $\bar{\nu}(\eta)=\exp(-\eta^2/2)$
   by the following function:
\begin{equation}
\bar{\nu}(\eta)=\mbox{sech}^2(b\eta),
\label{51}
\end{equation}
where the shape parameter $b$ is determined by the requirement that the total mass of the disc does not
change, namely:
\begin{equation}
\int^\infty_0\exp(-\eta^2/2) d\eta=
\int^\infty_0\mbox{sech}^2(b\eta) d\eta.
\label{52}
\end{equation}
The result is:
\begin{equation}
b=\sqrt{2/\pi}.
\label{53}
\end{equation}
The comparison of true and model number density profiles is presented in Fig. 1 and demonstrates a close correspondence between them. Furthermore, it will be shown below that the results derived by employing that profile are weakly distinguishable from the WKB results obtained for the true exponential distribution. In fact, notwithstanding the use of the terms true and model profiles, such a change of the equilibrium profile of the number density (that is determined by the axial momentum balance equation) may actually represent some true equilibrium that is obtained from a slightly different gravitational potential [see \cite{Spitzer 1942}, where a similar density profile has been obtained as an exact solution for flat disc-galaxies whose disc mass content is larger than the mass of the central object].

\begin{figure*}
\includegraphics[width=120mm]{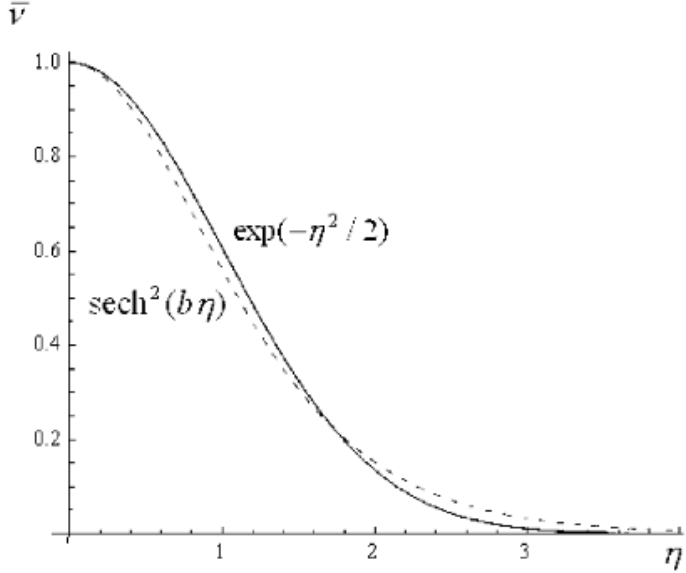}
  \caption{The comparison of true, $\bar{\nu}=\exp(-\eta^2/2)$  (solid) and model, $\bar{\nu}=\mbox{sech}^2(b\eta)$
  (dashed) number density profiles; $b=\sqrt{2/\pi}$.
  }
\label{Fig. 1}
\end{figure*}


Instead of Eq. (\ref{50}) for the perturbed magnetic field it is more convenient to consider the
equation for the perturbed velocity. Thus, substituting (\ref{51}) - (\ref{53}) into (\ref{45}) -
(\ref{48}), yields the following ordinary differential equations for  $\hat{v}_r$ and $\hat{v}_\theta$:
\begin{equation}
\frac{1}{\bar{\nu} (\eta)}
\frac{d^2}{d\eta^2}
 \left(\frac{1}{\bar{\nu}(\eta)}
 \frac{d^2\hat{v}_{r,\theta}}{d\eta^2}\right)
+(3+2\lambda^2)\bar{\beta}_z
\frac{1}{\bar{\nu}(\eta)}\frac{d^2
\hat{v}_{r,\theta}}{d\eta^2}
+\lambda^2( \lambda^2-1)\bar{\beta}_z^2  \hat{v}_{r,\theta}=0.
\label{54}
\end{equation}

The next step now is to introduce a new independent variable $\xi=\mbox{tanh}(b\eta)$, such that $-1\leq\xi\leq 1$.
Doing that, a simpler equation emerges, which may be cast into the following form:
\begin{equation}
(L+K^-)(L+K^+)\hat{v}_{\theta}=0,\,\,\,
\label{55}
\end{equation}
where $L$  is the Legendre operator of second order:
$$
L=\frac{d}{d\xi}[(1-\xi^2) \frac{d}{d\xi}],\,\,\,\,
K^\pm=\frac{\bar{\beta}_z}{2b^2}[3+2\lambda^2\pm\sqrt{9+16\lambda^2}].\,\,\,\,
\,\,\,\,\,\,\,\,\,\,\,\,\,\,\,\,\,\,\,\,\,\,\,\,\,\,\,\,\,\,\,\,\,\,\,\,\,\,\,\,\,\,\,\,\,\,\,\,\,\,\,\,\,\,
\,\,\,\,\,\,\,\,\,\,\,\,\,\,\,\,\,\,\,\,\,\,\,\,\,\,\,\,\,\,\,\,\,\,\,\,\,\,\,\,\,\,\,\,\,\,\,\,\,\,\,\,\,\,
\,\,\,\,\,\,\,\,\,\,\,\,\,\,\,\,\,\,\,\,\,\,\,\,\,\,\,\,\,\,\,\,\,\,\,\,\,\,\,\,\,\,\,\,\,\,\,\,\,\,\,\,\,\,
\,\,\,\,\,\,\,\,\,\,\,\,\,\,\,\,\,\,\,\,\,\,\,\,\,\,\,\,\,\,\,\,\,\,\,\,\,\,\,\,\,\,\,\,\,\,\,\,\,\,\,\,\,\,
\,\,\,\,\,\,\,\,\,\,\,\,\,\,\,\,\,\,\,\,\,\,\,\,\,\,\,\,\,\,\,\,\,\,\,\,\,\,\,\,\,\,\,\,\,\,\,\,\,\,\,\,\,\,
$$
Imposing now the zero boundary conditions  of $\hat{b}_\theta$  at $\eta\to \infty$ leads to the requirement that the solution of Eq.
(\ref{55}) for $\hat{v}_{\theta}$ diverges polynomially at most when $\eta\to \infty$. It is concluded therefore that $\hat{v}_{\theta}$ is proportional to the Legendre polynomials $P_k(\xi)$
 which are the solutions of the following equation:
\begin{equation}
\frac{d}{d\xi}\left((1-\xi^2) \frac{d P_k}{d\xi}\right)
+K^\pm P_k,\,\,\, K^\pm=k(k+1), \,\,\,k=1,2,....
\label{56}
\end{equation}
Thus, the eigenvalues $\lambda^\pm$ are now determined by the dispersion relation
\begin{equation}
K^\pm\equiv \frac{\bar{\beta}_z(r)}{2b^2}[3+2\lambda^2\pm\sqrt{9+16\lambda^2}]=k(k+1).
\label{57}
\end{equation}
Employing now the recursive relations for the Legendre polinomyals:
$$
 (1-\xi^2) \frac{d P_k}{d\xi}=\frac{k(k+1)}{2k+1} [P_{k-1}(\xi) - P_{k+1}(\xi)],
\,\,\,\,\,\,\,\,\,\,\,\,\,\,\,\,\,\,\,\,\,\,\,\,\,\,\,\,\,\,\,\,\,\,\,\,\,\,\,\,\,\,\,\,\,\,\,\,\,\,\,\,\,\,
\,\,\,\,\,\,\,\,\,\,\,\,\,\,\,\,\,\,\,\,\,\,\,\,\,\,\,\,\,\,\,\,\,\,\,\,\,\,\,\,\,\,\,\,\,\,\,\,\,\,\,\,\,\,
\,\,\,\,\,\,\,\,\,\,\,\,\,\,\,\,\,\,\,\,\,\,\,\,\,\,\,\,\,\,\,\,\,\,\,\,\,\,\,\,\,\,\,\,\,\,\,\,\,\,\,\,\,\,
\,\,\,\,\,\,\,\,\,\,\,\,\,\,\,\,\,\,\,\,\,\,\,\,\,\,\,\,\,\,\,\,\,\,\,\,\,\,\,\,\,\,\,\,\,\,\,\,\,\,\,\,\,\,
\,\,\,\,\,\,\,\,\,\,\,\,\,\,\,\,\,\,\,\,\,\,\,\,\,\,\,\,\,\,\,\,\,\,\,\,\,\,\,\,\,\,\,\,\,\,\,\,\,\,\,\,\,\,
$$
 setting  the arbitrary amplitude of $\hat{v}_{\theta}$  to unity, and using the linear
equations (\ref{45}) - (\ref{48}) result in the following expressions for the  eigenfunctions that are determined up to an
arbitrary radius dependent amplitude factor:
$$
\hat{v}_{r}^\pm =i\lambda^\pm_a \left(\frac{1}{2}
+\frac{3}{2}\frac{1}{\hat{\beta}_z}\frac{\hat{\beta}_z -1}{(\lambda^\pm_a)^2}\right) P_k(\xi),\,\,\
\hat{v}_{\theta}^\pm = P_k(\xi),\,\,\,
\,\,\,\,\,\,\,\,\,\,\,\,\,\,\,\,\,\,\,\,\,\,\,\,\,\,\,\,\,\,\,\,\,\,\,\,\,\,\,\,\,\,\,\,\,\,\,\,\,\,\,\,\,\,
\,\,\,\,\,\,\,\,\,\,\,\,\,\,\,\,\,\,\,\,\,\,\,\,\,\,\,\,\,\,\,\,\,\,\,\,\,\,\,\,\,\,\,\,\,\,\,\,\,\,\,\,\,\,
\,\,\,\,\,\,\,\,\,\,\,\,\,\,\,\,\,\,\,\,\,\,\,\,\,\,\,\,\,\,\,\,\,\,\,\,\,\,\,\,\,\,\,\,\,\,\,\,\,\,\,\,\,\,
\,\,\,\,\,\,\,\,\,\,\,\,\,\,\,\,\,\,\,\,\,\,\,\,\,\,\,\,\,\,\,\,\,\,\,\,\,\,\,\,\,\,\,\,\,\,\,\,\,\,\,\,\,\,
$$
$$
\hat{b}_r^\pm =\frac{k(k+1)}{2k+1}\frac{b}{6} [(\lambda^\pm_a-1) (\lambda^\pm_a+1) \hat{\beta}_z-3] \, [P_{k-1}(\xi) -
P_{k+1}(\xi)],
\,\,\,
\,\,\,\,\,\,\,\,\,\,\,\,\,\,\,\,\,\,\,\,\,\,\,\,\,\,\,\,\,\,\,\,\,\,\,\,\,\,\,\,\,\,\,\,\,\,\,\,\,\,\,
\,\,\,\,\,\,\,\,\,\,\,\,\,\,\,\,\,\,\,\,\,\,\,\,\,\,\,\,\,\,\,\,\,\,\,\,\,\,\,\,\,\,\,\,\,\,\,\,\,\,\,\,\,\,
\,\,\,\,\,\,\,\,\,\,\,\,\,\,\,\,\,\,\,\,\,\,\,\,\,\,\,\,\,\,\,\,\,\,\,\,\,\,\,\,\,\,\,\,\,\,\,\,\,\,\,
\,\,\,\,\,\,\,\,\,\,\,\,\,\,\,\,\,\,\,\,\,\,\,\,\,\,\,\,\,\,\,\,\,\,\,\,\,\,\,\,\,\,\,\,\,\,\,\,\,\,\,\,\,\,
$$
\begin{equation}
\hat{b}_{\theta}^\pm =\frac{k(k+1)}{2k+1}\frac{b}{4} \frac{(1-\lambda^\pm_a) (\lambda^\pm_a+1) \hat{\beta}_z-1 }{i\lambda^\pm_a
}\,[P_{k-1}(\xi) - P_{k+1}(\xi)],
\label{58}
\end{equation}
where $k=1,2,...$
plays the role of axial wave number, $\hat{b}_{\theta}^\pm=0$  for $\xi=\pm 1$, since $P_k(1)=1$  and $P_k(-1)=(-1)^k$   for all $k$.

Turning back to the  dispersion relation (58), it may be written as follows:
\begin{equation}
(\lambda^\pm)^4 \hat{\beta}_z^2- (\lambda^\pm)^2 \hat{\beta}_z(\hat{\beta}_z+6)
+9(1-\hat{\beta}_z)=0,\,\,\,\, \hat{\beta}_z=\frac{\bar{\beta}_z }{\bar{\beta}_{cr}^{(k)} }.
\label{59}
\end{equation}
It is thus obvious that the $k$-th mode is destabilized when the beta value crosses from bellow the threshold that is given by:
\begin{equation}
\bar{\beta}_{cr}^{(k)}=\frac{2}{3\pi}k(k+1).
\label{60}
\end{equation}
As a result, a universal (for all values of $\bar{\beta}_z(r)$ and  $k$) criterion for instability emerges which reads:  $\hat{\beta}_z(r)>1$.
Written in terms of the scaled plasma beta $\hat{\beta}_z $  the dispersion relation (\ref{59}) has the following
solutions for the eigenvalues of the Alfv\'en-Coriolis modes (see Fig. 1):
\begin{equation}
\lambda^\pm=\sqrt{
\frac{
\hat{\beta}_z +6\pm\sqrt{
     (\hat{\beta}_z+6)^2-36(1-\hat{\beta}_z)
                        }
}{2\hat{\beta}_z}.
}
\label{61}
\end{equation}
The two eigenvalues, $\lambda^+$   and  $\lambda^-$, represent fast and slow Alfv\'en-Coriolis waves. While
the fast Alfv\'en-Coriolis modes are always stable, the number of unstable slow modes signified by
 the plasma beta.
The eigenvalues of the slow Alfv\'en-Coriolis modes, $\lambda^-$  , are given therefore by:
$$
\lambda^-=\Lambda^-_a=\pm \sqrt{
\frac{
\hat{\beta}_z +6-\sqrt{
     (\hat{\beta}_z+6)^2-36(1-\hat{\beta}_z)
                        }
}{2\hat{\beta}_z}
},     \,\mbox{Im}\{\lambda^-)=0\,\,\,\,\mbox{for}\,\,\,\hat{\beta}_z\leq 1,
\,\,\,\,\,\,\,\,\,\,\,\,\,\,\,\,\,\,\,\,\,\,\,\,\,\,\,\,\,\,\,\,\,\,\,\,\,\,\,\,\,\,\,\,\,\,\,\,\,\,\,\,\,\,\,\,\,\,\,\,\,\,\,\,\,\,\,\,\,\,\,\,\,\,\
\,\,\,\,\,\,\,\,\,\,\,\,\,\,\,\,\,\,\,\,\,\,\,\,\,\,\,\,\,\,\,\,\,\,\,\,\,\,\,\,\,\,\,\,\,\,\,\,\,\,\,\,\,\,\,\,\,\,\,\,\,\,\,\,\,\,\,\,\,\,\,\,\,\,\
$$
\begin{equation}
\lambda^-=i\Gamma^-_a=\pm i \sqrt{
\frac{
\hat{\sqrt{
     (\hat{\beta}_z+6)^2+36(\hat{\beta}_z-1)-\beta}_z -6
                        }
}{2\hat{\beta}_z}
},     \,\mbox{Re}\{\lambda^-)=0\,\,\,\,\mbox{for}\,\,\,\hat{\beta}_z> 1.
\label{62}
\end{equation}
The eigenvalues are imaginary, and the system is spectrally unstable if $\hat{\beta}_z >1$, and real for
$\hat{\beta}_z <1$  in which case the system is stable. In particular, the minimal critical unscaled plasma
beta that is needed for instability is determined by the first unstable slow Alfv\'en-Coriolis mode, $k=1$, and is given by $\bar{\beta}_{cr}^{(1)}=0.42$.
The unstable modes are the well known MRIs  and from Eq. (\ref{60}) it is easy to
calculate how many of them are excited for a given value of the plasma beta.
 Thus, there are $k$ unstable modes for $\bar{\beta}_{cr}^{(1)}\leq\bar{\beta}_z(r)\leq\bar{\beta}_{cr}^{(k+1)}$.
 In particular, Eq. (\ref{60}) yields approximately the square-root law for the number of the unstable modes vs plasma beta:
\begin{equation}
k<\sqrt{3\pi\bar{\beta}_z(r)/2}.
\label{63}
\end{equation}
Relation (\ref{60}) or its simplified version (\ref{63}) is significant for the consequent modeling of non-linear development of the instability. It is finally emphasized that the stability criterion as well as the number of unstable modes depend on the radius. Thus, different areas within the disc may be characterized by different stability properties as well as different number of unbstable modes.

The family of the fast Alfv\'en-Coriolis modes, $\lambda^+(\hat{\beta}_z)$, is characterized by the frequencies that are
much larger than the Keplerian frequency for small values of the scaled plasma beta (large values of axial wave number or
small plasma beta) and tends to the Keplerian value at large plasma beta values:
\begin{equation}
\lambda^+=\Lambda^+_a=\pm \sqrt{
\frac{
\hat{\beta}_z +6+\sqrt{
     (\hat{\beta}_z+6)^2-36(1-\hat{\beta}_z)
                        }
}{2\hat{\beta}_z}
},     \,\mbox{Im}\{\lambda^+)=0.
\label{64}
\end{equation}

Expressed in terms of the scaled plasma beta, a single figure depicts all possible stable as well as unstable
modes. This is shown in Fig. 2. The maximal growth rate for the unstable modes is achieved around
$\hat{\beta}_z \approx3$, which for a given plasma beta value determines the  axial wave number of the
fastest growing modes.


\begin{figure*}
\includegraphics[width=120mm]{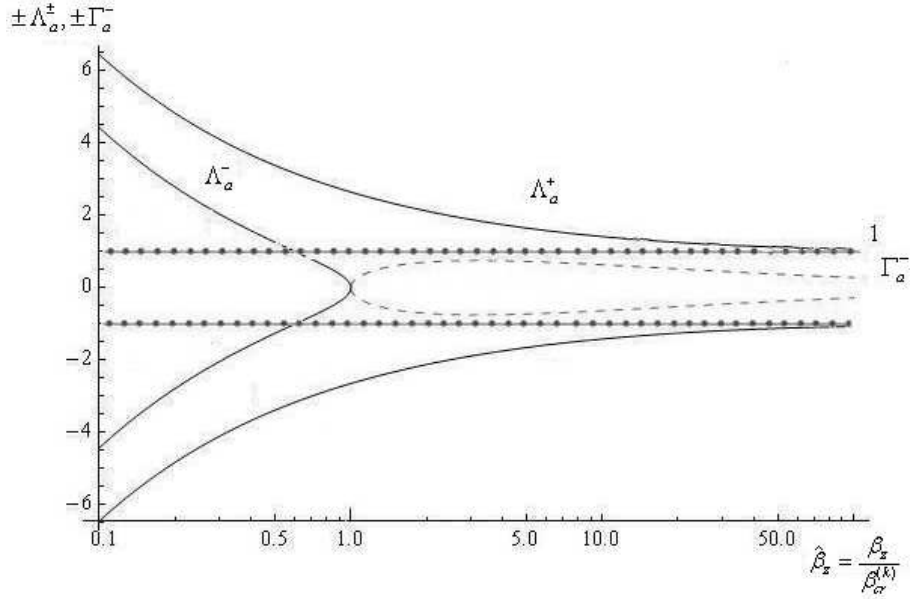}
  \caption{Growth rates $\pm\Gamma^-_a$  (dashed lines) for the unstable Alfv\'en-Coriolis (MRI) modes and frequencies
  $\pm\Lambda^\pm_a$  (solid lines) for the Alfv\'en-Coriolis oscillations vs universal scaled plasma beta,
  $\hat{\beta}_z =\bar{\beta}_z(r)/ =\bar{\beta}_{cr}^{(k)}$, $\bar{\beta}_{cr}^{(k)}=2k(k+1)/(3\pi)$ for the model number density
  $\bar{\nu}=\mbox{sech}^2(\sqrt{2/\pi}\eta)$;
$k=1,2,\dots$  is the axial wave number. Meshed straight-line asymptotes at $\hat{\beta}_z \gg 1$  are the scaled
Keplerian frequencies, $\pm\Lambda^\pm_a=\pm 1$.
  }
\label{Fig. 2}
\end{figure*}

Frequencies and growth rates for the first five fast and slow Alfv\'en-Coriolis  modes are presented in Fig. 3
(in terms of the unscaled beta, $\bar{\beta}_z $ ). The corresponding  growth rates $\Gamma^-_a$
obtained by Liverts \& Mond (2009) by employing the WKB approximation are
weakly distinguishable from  those in Fig. 3 and therefore are not shown on the figure. In particular, the critical plasma beta
 $\bar{\beta}^{(1)}_{cr} $  found by \cite{Liverts and Mond 2009} (after correction by factor 2 due to the different
 definition of the plasma beta there) is $0.417$, which is practically the same as the one obtained in the
 current work.

\begin{figure*}
\includegraphics[width=175mm]{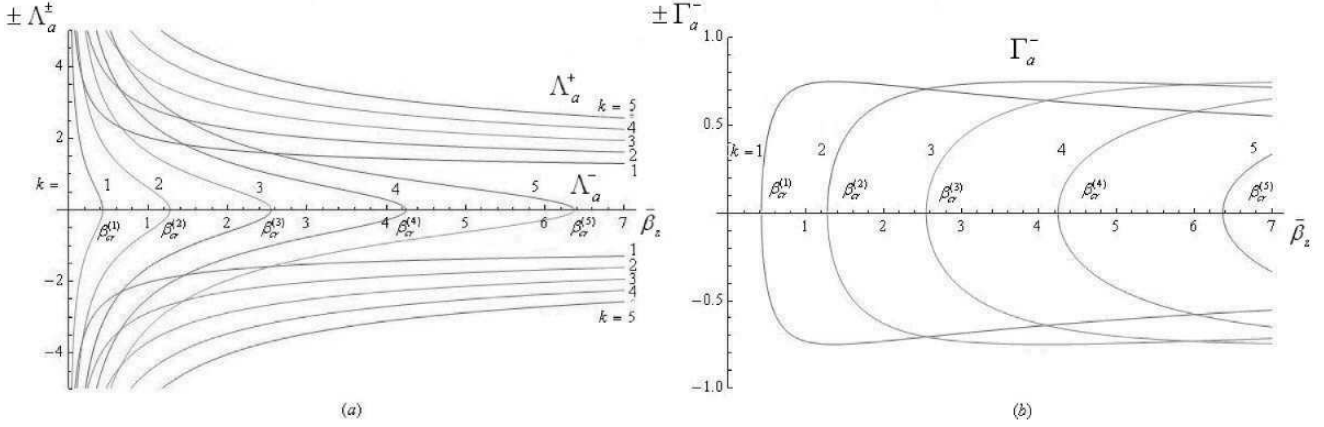}
 \caption{
  Frequencies $\pm\Lambda^\pm_a$  for the first five fast and slow Alfv\'en-Coriolis modes and growth rates $\pm\Gamma^-_a$
    for the unstable (MRI) slow Alfv\'en-Coriolis  modes vs unscaled plasma beta $\bar{\beta}_z$  for the model number density
   $\bar{\nu}=\mbox{sech}^2(\sqrt{2/\pi}\eta)$;
$\bar{\beta}_{cr}^{(k)}=2k(k+1)/(3\pi)$  is the critical plasma beta, $k=1,2,3,4,5$  is the axial wave number.
  }
\label{Fig. 3}
\end{figure*}

An illustration of the perturbed toroidal magnetic field is presented in Fig. 4. The perturbations are indeed localized within the effective height of the disc which weakly depends on the axial wave number. This corresponds to a finite distance between the turning points -- the natural characteristics of the problem solution in the WKB approximation [\cite{Liverts and Mond 2009}]. To explicitly demonstrate the finite size of the region of perturbation's location, the differential equation for the perturbed magnetic field  $\hat{b}_\theta$ in the self - similar variable $\eta$  is written out:
 \begin{equation}
 \frac{d^2 \hat{b}_\theta}{ d\eta^2}+2 b\xi \frac{d \hat{b}_\theta}{ d\eta}+b^2K^{\pm}(1-\xi^2)\hat{b}_\theta=0,\ \ \
\xi=\mbox{tanh}(b\eta),\ \ \ b=\sqrt{\frac{2}{\pi}},\ \ \ K^{\pm}=k(k+1),\ \ \ k=1,2,\dots,
\label{b theta eta}
\end{equation}
  which is subject to zero boundary conditions at infinity. Equation (\ref{b theta eta}) is derived by using differential equation (\ref{55}) for $\hat{v}_\theta$  and a fact that $\hat{b}_\theta$ up to a constant coefficient of proportionality is equal to $d\hat{v}_\theta/d\eta$.
Transforming as in WKB approximation  the dependent variable   $\hat{b}_\theta=Q(\eta)\int\mu(\eta)d\eta$, and setting  $\mu(\eta)=-\xi=-\mbox{tanh}(b\eta)$ in order to eliminate terms with  first order derivatives  yield:
 \begin{equation}
 \frac{d^2 Q}{ d\eta^2}+ \varkappa^2(\eta)Q=0,\ \ \ \varkappa^2(\eta)=b^2[K^{\pm}(1-\xi^2)-1].
\label{Q eta}
\end{equation}
Then the turning points are   determined from the relation $\varkappa^2(\eta)=0$, and may be
found explicitly
substituting   $1-\xi^2$ by
$\mbox{sech}^2(b\eta)$:
 \begin{equation}
\eta_*=b^{-1} \mbox{arcsech}\frac{1}{\sqrt{k(k+1)}},\ \ \ k=1,2,\dots.
\label{turning points eta}
\end{equation}
\begin{figure*}
\includegraphics[width=175mm]{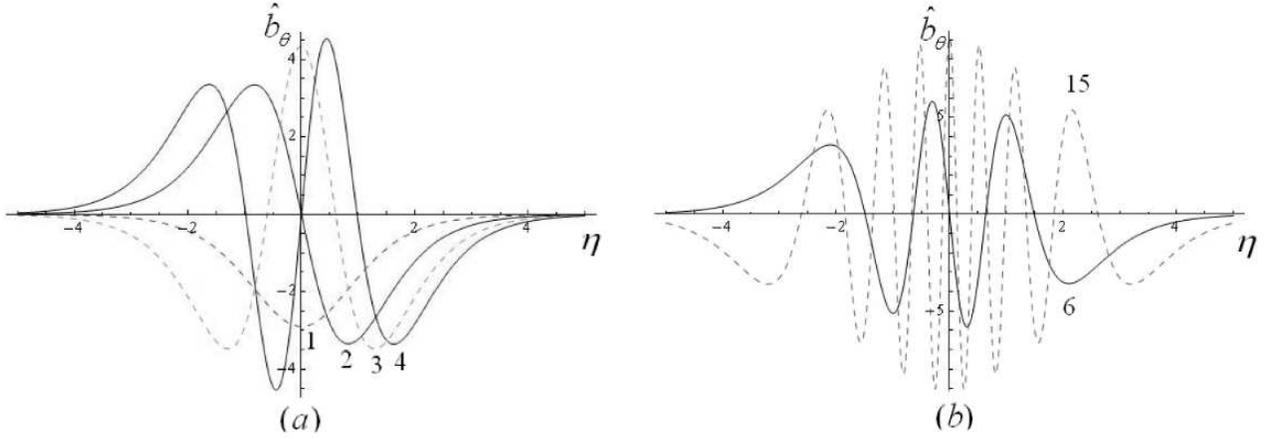}
 \caption{
 The toroidal component of the perturbed magnetic field  $\hat{b}_\theta$ vs self-similar axial variable $\eta=z/H(r)$  for the Legendre polynomials with the axial wave numbers (a)  $k=1, 2, 3,4$ and (b)   $k=6,\,15$.
All curves are calculated for the fixed value of  $\hat{\beta}_z=1.5$ ($\hat{\beta}_z=\bar{\beta}_z/\bar{\beta}_{cr}^{(k)}$ ,  $\bar{\beta}_{cr}^{(k)}=b^2k(k+1)/3$, dashed and solid  curves correspond to the odd  and even  $k$, respectively).
  }
\label{Fig. 4}
\end{figure*}
The results for several axial wave numbers are presented in Table 3. Although the oscillation's region unboundedly expanded with rising axial wave numbers, this occurs very slowly and the region has relatively low sizes  even at high values of $k$. Note that the unbounded growth which corresponds to high wave numbers should be suppressed by dissipative effects neglected in the present analysis.

 \begin{table*}
 \centering
 \begin{minipage}{140mm}
\caption{Distances of the turning points from the midplane vs axial wave number.
}
\begin{tabular}{@{}cccccccccc@{}}
$k$
& $1$ & $2$ & $3$ & $4$
& $5$ & $6$ & $15$ & $35$ &  $155$ 
    \\
      \hline
        $\eta_*$
       & $1.1$  &  $1.9$  &  $2.4$  & $2.7$
              &  $3.0$  &   $3.2$  &   $4.3$
              &  $5.3$  &   $7.2$  
                          \\
        \end{tabular}
\end{minipage}
\end{table*}

In order to compare the current results to well known results for infinite homogeneous cylinders \cite{Balbus and Hawley 1991} the growth rate of the unstable modes is depicted in Fig. 5 as a function of the axial wave number  $K$, where
 \begin{equation}
K=k \frac{L_a}{ H\sqrt{2}}\equiv \frac{k}{\sqrt{2\bar{\beta}_z}},
\label{65}
\end{equation}
$L_a=V_a/\Omega$ is the Alfv\'en length scale, $z=H\sqrt{2}$ is the effective height of the diffused disc at which the equilibrium number density, $\bar{n}\sim\exp[-z^2/(2H^2)]$, falls by factor $e^{-1}$.
 The effective  wave number $K$ is the discrete thin-disc analog of the continuous wave number for  infinite cylindrical discs. For fixed value of the local plasma beta,  $\bar{\beta}_z=0.41,\,\,0.5,\,\,1.5,\,\,2.5,\,500$,  the discrete set of the points in the plane $\{K,\Gamma^{-}_a\}$ is presented by one of the interpolating curves $1, 2, 3, 4, 5$, respectively. The number of the discrete  points on each interpolating curves  corresponds to the admissible values of the Alfv\'en-Coriolis  mode number $k=1,2,3,...$ for which $\bar{\beta}_{cr}^{(1)}\leq\bar{\beta}_{cr}^{(k)}\leq\bar{\beta}_z(r)$.
 Also, the range of unstable $k$-values is widening as the value of $\bar{\beta}_z$  is increased.
 In particular, at large plasma beta the corresponding set of  the points due to their large number (curve 5) should tend to the continuous curve for infinite cylinder geometry in \cite{Balbus and Hawley 1991}. For finite plasma beta there is a discrete number of points on each curves in Fig. 5, where the left bound corresponds to the  first Alfv\'en-Coriolis mode, $k=1$.
 Thus, for beta values close to  $\bar{\beta}_{cr}^{(1)}$, the number of unstable modes is small and the disk stability properties significantly deviate from those predicted by the infinite cylinder model.

\begin{figure*}
\includegraphics[width=115mm]{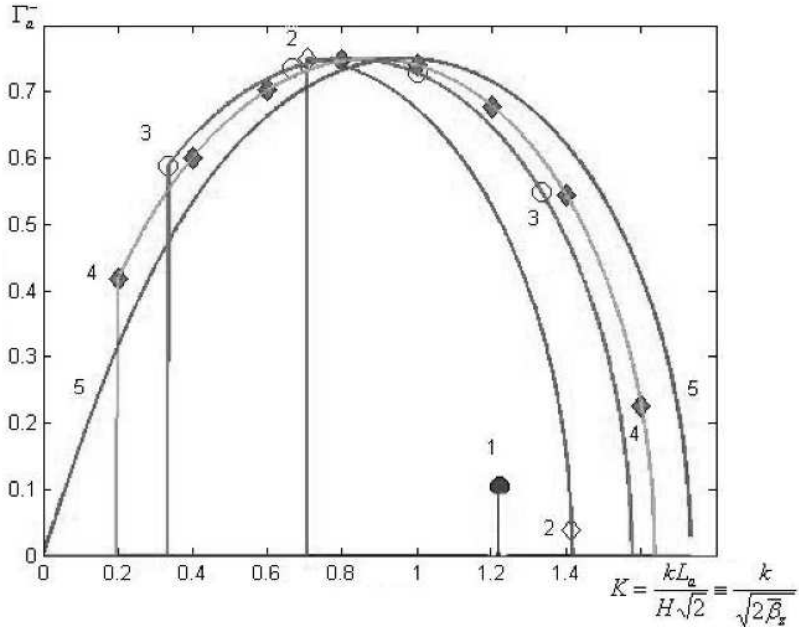}
 \caption{
Growth rates $\Gamma^-_a$   for the slow Alfv\'en-Coriolis modes vs effective wave number $K=\frac{k L_a}{H \sqrt{2}}\equiv\frac{k}{\sqrt{2\bar{\beta}_z}}$,
 $k=1,2,3,...$  is the number of the Alfv\'en-Coriolis modes, calculated for the model number density  $\bar{\nu}=\mbox{sech}(b\eta)$. Interpolating curves $1, 2, 3, 4, 5$ correspond to
 $\bar{\beta}_z=0.41,\,\,0.71,\,\,1.5,\,\,2.5,\,500$, respectively.
  }
\label{Fig. 4}
\end{figure*}



\subsection{Linear stability problem for magnetosonic modes}
For simplicity it is assumed now that the poloidal component of the equilibrium magnetic field is zero and hence   $\bar{\beta}_z=\infty$. Thus,
substituting the exponential anzatz (\ref{43}) into Eqs. (\ref{41}) - (\ref{42}) for the magnetosonic modes, and assuming that their frequencies are different from the eigenvalues of inertial-Coriolis waves  yield:
\begin{equation}
-i\lambda \hat{v}_z
+
\frac{d}{d \eta }[\frac{\hat{\nu}}{\bar{\nu}(\eta)}]
=
0,
\label{66}
\end{equation}
\begin{equation}
-i\lambda  \frac{\hat{\nu}}{\bar{\nu}(\eta)}
+\frac{d\hat{v}_z}{d \eta }+\frac{1}{\bar{\nu}(\eta)}\frac{d \bar{\nu}(\eta)}{d \eta}\hat{v}_z=0.
\label{67}
\end{equation}
Since the effect of the magnetic field  is dropped from the linear system (\ref{66}) -(\ref{67}), the latter actually
describes sound waves that propagate vertically in the non-homogeneous disc. Eliminating the  perturbed number density from
 (\ref{66}) -(\ref{67}) in the resulting equation yields a second order differential equation for the perturbed axial velocity:
 \begin{equation}
\frac{d^2 \hat{v}_z}{d \eta^2 }+\frac{1}{\bar{\nu}(\eta)}\frac{d \bar{\nu}(\eta)}{d \eta }\frac{d \hat{v}_z}{d \eta }
+\big{\{}\lambda^2+\frac{d }{d \eta }\big[\frac{1}{\bar{\nu}(\eta)}\frac{d \bar{\nu}(\eta)}{d \eta }
\big]\big{\}}\hat{v}_z=0,
\label{68a}
\end{equation}
or substituting
the equilibrium number density, $\bar{\nu}(\eta)=\exp(-\eta^2/2)$,
\begin{equation}
\frac{d^2 \hat{v}_z}{d \eta^2 }-\eta\frac{d \hat{v}_z}{d \eta }+(\lambda^2-1)\hat{v}_z=0.
\label{68}
\end{equation}
Equation (\ref{68}) is the Hermite equation and the requirement that its solutions diverges polynomially
at most when $\eta\to\pm\infty$  determines the eigenvalues to be:
\begin{equation}
\lambda=\pm \Lambda_S=\pm\sqrt{m+1},
\label{69}
\end{equation}
while the eigenfunctions are given up to an amplitude factor depending on radius by:
\begin{equation}
\hat{v}_z=\mp i \sqrt{m+1}H_{m}(\eta),\,\,\, \hat{\nu}=e^{-\eta^2/2}H_{m+1}(\eta).
\label{70}
\end{equation}
Here $ H_{m}(\eta)$  ($m=1,2,…$ ) are the Hermite polynomials. The axial velocities are polynomially unbounded
functions at $\eta\to\pm \infty$.
Note that now the number of the Hermite polynomial $m$ plays the role of the axial wave number for the
magnetosonic mode that is different from the Alfv\'en-Coriolis  modes with the number of the Legendre polynomial  as the axial
wave number. Formally, the above family of the solutions of the eigenvalue problem should be completed by the
following degenerate steady-state solution:
\begin{equation}
\hat{\nu}=e^{-\eta^2/2},\,\,\, \hat{v}_z=0.
\label{71}
\end{equation}
However, the latter may be considered as a modification of the unperturbed equilibrium solution.

\section{DYNAMICAL EQUATIONS FOR PERTURBED THIN DISCS: TYPE II EQUILIBRIA }
Consider now the thin disc problem under the effect of a dominant toroidal component of the equilibrium magnetic fields (see Tables 1 and 2). The solutions of such problem in the pure hydrodynamic case of zero magnetic field for adiabatic thin discs have been obtained and discussed in \cite{Shtemler et al. 2010} and are extended here for locally isothermal magnetized discs.

\subsection{The reduced nonlinear equations}
At this stage it is convenient to rescale all the physical variables, $f$, with the small parameter $\epsilon$, without splitting the solution to its equilibrium and perturbed constituents. Thus, in a most general way, the physical variables are given by:
\begin{equation}
f(r,\zeta,t) =\epsilon^{S'} \tilde{f}(r,\zeta,t).
\label{72}
\end{equation}
Here the auxiliary variables denoted by tilde are introduced, and the same orders,   $S'$, in  $\epsilon$  are assumed for the physical variables as for the disturbed variables. For the dominant toroidal equilibrium  magnetic field such choice is justified by the fact that any disturbed dependent variable is of the order of or larger than the corresponding equilibrium variable i.e.  $S'\leq\bar{S}$   (see Tables 1 and 2).
This imposes an additional limitation on the admissible time of perturbation growth.
Substituting  (\ref{72}) into (\ref{2}) -(\ref{6})   yields up to the terms of the higher order in  $\epsilon$    the following in-plane components of the momentum equation:
\begin{equation}
\frac{\partial \tilde{V}_r}{\partial t}
+\tilde{V}_r\frac{\partial \tilde{V}_r }{\partial r}
+\tilde{V}_z\frac{\partial \tilde{V}_r}{\partial \zeta}
-\frac{ \tilde{V}_\theta^2}{r}
=-\frac{d \bar{\Phi}}{d r},
\label{73}
\end{equation}
\begin{equation}
\frac{\partial \tilde{V}_\theta }{\partial t}
+\tilde{V}_r\frac{\partial \tilde{V}_\theta }{\partial r}
+\tilde{V}_z\frac{\partial \tilde{V}_\theta }{\partial \zeta}
+\frac{\tilde{V}_r \tilde{V}_\theta}{r}
=0,
\label{74}
\end{equation}
%
\begin{equation}
\frac{\partial \tilde{V}_z }{\partial t}
+\tilde{V}_r \frac{\partial \tilde{V}_z}{\partial r}
+\tilde{V}_z \frac{\partial \tilde{V}_z}{\partial \zeta}
=-\frac{\bar{c}^2_S}{\tilde{n}}\frac{\partial \tilde{n}}{\partial \zeta}
-\frac{\partial \bar{\phi}}{\partial \zeta}
-\frac{1}{2\beta\tilde{n}}
\frac{\partial (\tilde{B}_\theta^2+\tilde{B}_r^2)}{\partial \zeta },
\label{75}
\end{equation}
\begin{equation}
\frac{\partial  \tilde{n}}{\partial t}
+\frac{1}{r} \frac{\partial (r\tilde{n} \tilde{V}_r )}{\partial r }
+ \frac{\partial (\tilde{n} \tilde{V}_z )}{\partial \zeta}=0,
\label{76}
\end{equation}
\begin{equation}
\frac{\partial \tilde{B}_r}{\partial t}
-\frac{\partial \tilde{E}_\theta }{\partial \zeta}=0,
\label{77}
\end{equation}
\begin{equation}
\frac{\partial \tilde{B}_\theta}{\partial t}
+\frac{\partial \tilde{E}_r}{\partial \zeta}
-\frac{\partial \tilde{E}_z}{\partial r}
=0,
\label{78}
\end{equation}
%
%
\begin{equation}
\frac{\partial \tilde{B}_z}{\partial t}
+\frac{1}{r}\frac{\partial (r\tilde{E}_\theta)}{\partial r}
=0.
\label{79}
\end{equation}
Here
\begin{equation}
\tilde{E}_r=\tilde{V}_z\tilde{B}_\theta - \tilde{V}_\theta\tilde{B}_z,\,\,\,
\tilde{E}_\theta =\tilde{V}_r\tilde{B}_z - \tilde{V}_z\tilde{B}_r ,\,\,\,
\tilde{E}_z= \tilde{V}_\theta\tilde{B}_r - \tilde{V}_r\tilde{B}_\theta.
\label{80}
\end{equation}
The system (\ref{73})- (\ref{80}) is subject the boundary conditions  at infinity  of  least possible divergence for the axial velocity and the vanishing of the in-plane magnetic-field
components.

\subsection{The linearized problem }
We start by linearizing the MHD equations (\ref{73}) - (\ref{80})  about the steady-state equilibrium solution (\ref{12}) - (\ref{15}). Substituting the decomposition of the total disturbed variables (\ref{18}) into Eqs.  (\ref{73}) - (\ref{80}), yields the following linear set of equations for  axisymmetric perturbations:
\begin{equation}
\frac{\partial V_r'}{\partial t}
-2\bar{\Omega}(r)V_\theta'
=0,
\label{81}
\end{equation}
\begin{equation}
\frac{\partial V_\theta' }{\partial t}
+\frac{1}{2}\bar{\Omega}( r)V_r'
=0,
\label{82}
\end{equation}
and
\begin{equation}
\frac{\partial  V_z'}{\partial t}
+\bar{c}_S^2(r) \frac{\partial}{\partial \zeta }(\frac{n'}{\bar{n}})
+\frac{1}{\beta}\frac{\bar{B}_\theta(r)} {\bar{n}}
\frac{\partial B_\theta'}{\partial \zeta }=0,
\label{83}
\end{equation}
\begin{equation}
\frac{\partial n'}{\partial t}
+\frac{\partial (\bar{n}V_z')}{\partial \zeta }=-\frac{1}{r}\frac{\partial (r\bar{n}V_r')}{\partial r },
\label{84}
\end{equation}
%
%
\begin{equation}
\frac{\partial B_r'}{\partial t}
=\bar{B}_z(r)\frac{\partial V_r'}{\partial \zeta},
\label{85}
\end{equation}
\begin{equation}
\frac{\partial B_\theta'}{\partial t}
+\bar{B}_\theta(r)\frac{\partial V_z'}{\partial \zeta}=
r\frac{d \bar{\Omega}}{d r}B_r'
+ \bar{B}_z(r) \frac{\partial V_\theta'}{\partial \zeta }
- \frac{\partial [\bar{B}_\theta(r)V_r' ]}{\partial r }.\,\,\,
\label{86}
\end{equation}
As for the equilibria type I, the equation for the perturbed magnetic field, $B_r'$,  is obtained by differentiating by   the equation for the magnetic flux function, $\Psi'$, while the corresponding
equation for the perturbed axial magnetic field,  $B_z'$, decouples from the rest of the equations, and drops out altogether from the governing system (\ref{81})-(\ref{86}). As in the case of comparable axial and toroidal magnetic field components (type I equilibria), also the reduced linear system (\ref{81})-(\ref{86}) is decoupled into two linear sub-systems that describe now the dynamics of the inertial-Coriolis (IC), $\{ V_r',v_\theta'\}$, and magnetosonic (MS), $\{V_z',\,n',B_r'\,,B_\theta',\,B_z'\}$, modes. A notable difference between the two cases however is that, while for the type I equilibria the radial derivatives drop out altogether from the reduced nonlinear system of equations, in the current case, some of the radial derivatives do survive the asymptotic procedure, due to the relative smallness of the axial magnetic field component. As will be shown later on, this difference turns out to be quite significant as the radial derivatives are responsible for the non modal algebraic growth of the initially small perturbations.

Back to the modes of wave propagation, the IC waves are described by Eqs.  (\ref{81}) and (\ref{82}) while the MS modes are given by the solution of the homogeneous parts of Eqs.  (\ref{83})-(\ref{86}). Also, the perturbed radial magnetic field,  $B_r'$, is completely determined by the perturbed radial velocity,  $V_r'$, and hence Eq. (\ref{85}) is separated from the rest of the magnetosonic sub-system. The right hand sides of Eqs. (\ref{84}) and (\ref{86})  result in the (resonant as well as nonresonant) forcing of the MS waves by the IC modes. As will be seen later on, this occurs due to the rotational shear.

Before actually solving the above linearized system of equations it is noted that, guided by the steady-state solution, it is convenient to introduce the following new independent variables:
\begin{equation}
\theta=t,\,\,\, \rho=\int_0^r\frac{d r}{\bar{H}(r)},\,\,\,\eta=\frac{\zeta}{\bar{H}(r)},\,\,\, (\bar{H}(r) =\frac{\bar{c}_S(r)}{
\bar{\Omega}(r)}).
\label{88}
\end{equation}
Then derivatives in new and old variables are related as follows:
\begin{equation}
\frac{\partial }{\partial t} =\frac{\partial }{\partial \theta },\,\,\,
\frac{\partial}{\partial \zeta} = \frac{1 }{ \bar{H}}\frac{\partial }{\partial \eta },\,\,\,
\frac{\partial}{\partial r} =\frac{\partial}{\partial \rho} - \frac{1 }{ \bar{H}^2} \frac{d\bar{H}}{d \rho}
\eta\frac{\partial }{\partial \eta }.
\label{89}
\end{equation}
Equations (\ref{81})-(\ref{86}) may be rewritten by introducing the following scaled variables:
\begin{equation}
{\bf {v}}(\rho,\eta,\theta) = \frac{{\bf{V}}{'}}{\bar{c}_S(r)},\,\,\,
\nu(\rho,\eta,\theta) = \frac{n'}{\bar{N}(r)},\,\,\,
{\bf {b}}(\rho,\eta,\theta) = \frac{{\bf{B}}'}{\bar{B}_\theta(r)}.
\label{90}
\end{equation}
Below for convenience and with no confusion we revert to the notation $t$ for the new variable  $\theta$. This yields the following system of equations:
\begin{equation}
\frac{1}{\bar{\Omega}}\frac{\partial v_r}{\partial t}
-2 v_\theta =0,
\label{91}
\end{equation}
\begin{equation}
\frac{1}{\bar{\Omega}}  \frac{\partial v_\theta }{\partial t }
+\frac{1}{2} v_r =0,
\label{92}
\end{equation}
and
\begin{equation}
\frac{1}{\bar{\Omega}}  \frac{\partial v_z}{\partial t }
+\frac{\partial}{\partial \eta }[\frac{\nu}{\bar{\nu}(\eta)}]
+\frac{1}{ \bar{\beta}_\theta \bar{\nu}(\eta)}
\frac{\partial b_\theta}{\partial \eta }=0,
\label{93}
\end{equation}
\begin{equation}
\frac{1}{\bar{\Omega}}  \frac{\partial }{\partial t}[\frac{\nu}{\bar{\nu}(\eta)}]
+\frac{\partial v_z}{\partial \eta }-\eta v_z
=- \frac{\partial v_r}{\partial \rho }
-\bar{D}_{N} v_r
+\bar{D}_{\Omega} (\frac{\partial v_r}{\partial \eta }- \eta v_r)\eta,
\label{94}
\end{equation}
%
\begin{equation}
\frac{1}{\bar{\Omega}} \frac{\partial b_r}{\partial t}
=\frac{1}{\bar{S}(r)}\frac{\partial v_r}{\partial \eta},
\label{95}
\end{equation}
\begin{equation}
\frac{1}{\bar{\Omega}}   \frac{\partial b_\theta}{\partial t}
+\frac{\partial v_z}{\partial \eta}=-\frac{\partial v_r }{\partial \rho }
-\frac{3}{2}b_r
+ \frac{1}{\bar{S}(r)}\frac{\partial v_\theta}{\partial \eta }
-\bar{D}_{B}v_r+\bar{D}_{\Omega}\frac{\partial v_r }{\partial \eta
}\eta.\,\,
\label{96}
\end{equation}
Here $\bar{\beta}_\theta(\rho)$, $\bar{\beta}_z(\rho)$  and  $\bar{S}(r)$ are given by (\ref{36}); $\bar{\beta}_\theta(\rho)$  and $\bar{\beta}_z(\rho)$ are proportional to the characteristic lasma beta,  $\beta$, that is now based on the dimensional toroidal component of the equilibrium magnetic field as the characteristic scale, $B_*=B_\theta(r_*)$;
 $\bar{D}_N(\rho)$ , $\bar{D}_{\Omega}(\rho)$  and $\bar{D}_B(\rho)$  are the following logarithmic
derivatives:
\begin{equation}
\bar{D}_N=\frac{d \ln(\bar{c}_S r\bar{N})}{d \rho },\,\,\,
\bar{D}_{\Omega} =\frac{d \ln(\bar{c}_S /\bar{\Omega})}{d \rho }
\equiv\frac{d\ln\bar{H}}{d \rho},\,\,\,
\bar{D}_B =\frac{d \ln(\bar{c}_S \bar{B}_\theta)}{d \rho }.\,\,\,
\label{97}
\end{equation}

A single equation for the magnetosonic modes may be derived now by eliminating the toroidal component of the perturbed magnetic field as well as the perturbed number density from Eqs. (\ref{93}) - (\ref{96}). The result is:
\begin{equation}
\frac{1}{\bar{\Omega}^2}  \frac{\partial^2 v_z}{\partial t^2 }
-(1+\frac{1}{\bar{\beta}_\theta \bar{\nu}(\eta)})\frac{\partial^2 v_z }{\partial \eta^2 }
+\eta\frac{\partial v_z }{\partial \eta } + v_z
=\frac{1}{ \bar{\beta}_\theta
           \bar{\nu}(\eta)
          }
(\frac{3}{2}\frac{\partial b_r}{\partial \eta }
-\sqrt{\frac{\bar{\beta}_\theta}{ \bar{\beta}_z} }\frac{\partial^2 v_\theta}{\partial \eta^2 })
+(1+\frac{1}{\bar{\beta}_\theta \bar{\nu}(\eta)})
\frac{\partial^2 v_r}{\partial \eta \partial\rho}+L_z v_r,
\label{98}
\end{equation}
where $L_z$  is the linear ordinary differential  operator:
\begin{equation}
L_z v_r =
\bar{D}_N \frac{\partial  v_r}{\partial \eta }
-\bar{D}_{\Omega}[ (1+\frac{1}{\bar{\beta}_\theta \bar{\nu}(\eta)})
 \frac{\partial}{\partial \eta }( \eta \frac{\partial v_r}{\partial \eta })
-\frac{\partial  (\eta^2 v_r)}{\partial \eta }]
+\bar{D}_B \frac{1}{\bar{\beta}_\theta \bar{\nu}(\eta)} \frac{\partial v_r}{\partial \eta }.
\label{99}
\end{equation}

Note that in the important particular case of pure hydrodynamic discs the equilibrium and perturbed components of the magnetic field should be set to zero, i.e.
 $\bar{\beta}_\theta=\bar{\beta}_z=\infty$ and $b_r=b_\theta=0$, then Eqs. (\ref{96}) - (\ref{97}) with the properly resolved uncertainty in  $\bar{\beta}_\theta/\bar{\beta}_z$  are satisfied identically, and Eq. (\ref{98}) is replaced by
\begin{equation}
\frac{1}{\bar{\Omega}^2}  \frac{\partial^2 v_z}{\partial t^2 }
-\frac{\partial^2 v_z }{\partial \eta^2 }
+\eta\frac{\partial v_z }{\partial \eta } + v_z
=
\frac{\partial^2 v_r}{\partial \eta \partial\rho}+
\bar{D}_N \frac{\partial  v_r}{\partial \eta }
-\bar{D}_{\Omega}[
 \frac{\partial}{\partial \eta }( \eta \frac{\partial v_r}{\partial \eta })
-\frac{\partial  (\eta^2 v_r)}{\partial \eta }]
.
\label{100}
\end{equation}

The above problems formulated   for the magnetized and magnetic-field free discs are complemented by the  boundary condition  of  least possible divergence  at infinity for the axial velocity, and in the case of magnetized discs the vanishing conditions for the in-plane magnetic-field components should be also satisfied.

An important result is immediately seen from the above set of linearized equations: to leading order in  $\epsilon$: Eqs. (\ref{91})-(\ref{92}) decouple from the rest of the system in the thin disc approximation. Note additionally that Eq. (\ref{95}) for the radial magnetic field, explicitly determined by the radial velocity, may be solved separately from the rest Alfv\'en-mode sub-system. Thus, the sub-system (91)-(92) describes pure hydrodynamic in-plane inertia-Coriolis waves, and signifies the decoupling of the latter from the Alfv\'en waves. This occurs due to the negligibly small values of the projections on the disc plane of the pressure gradient and Lorentz force in the   momentum balance equations  (of the order of $\epsilon^2$) compared with the  inertial terms (of the order of   $\epsilon^0$) in the thin disc approximation. Thus, in the case of  poloidal-dominated equilibrium, the magnetic field is of order  $\epsilon^0$ which renders the Lorentz force of the same order of magnitude as the inertial terms, and thus couples the Lorentz and inertia forces. This results in the coupling of the Coriolis and Alfv\'en waves that eventually leads to the MRI. In contrast, the case of the toroidal-dominated  magnetic field inevitably results in the scaling   $\bar{B}_z\sim\epsilon$. This, as indicated above, means that the radial and azimuthal components of the Lorentz force are negligible with respect to the corresponding components of the inertial forces in the radial and azimuthal components of the momentum equation. As is already apparent, that decoupling between the inertial and Lorentz forces causes to a crucial deviation from the former case (i.e., with polidal-dominated equilibrium  magnetic field) and leads to the removal of the mechanism that is responsible to the MRI.

\subsection{ Linear stability problem.}
The perturbations may be presented  as follows
\begin{equation}
f(\rho,\eta,t)=\exp[-i\lambda\bar{\Omega}(\rho)t]\hat{f}(\rho,\eta),
\label{102}
\end{equation}
where $\lambda=\omega/\bar{\Omega}$  is the scaled frequency.

{\it{Inertia-Coriolis modes}}.
Substituting (\ref{102}) into (\ref{91}) - (\ref{92}) yields:
\begin{equation}
-i\lambda \hat{v}_r
-2 \hat{v}_\theta =0,
\label{103}
\end{equation}
\begin{equation}
-i\lambda \hat{v}_\theta
+\frac{1}{2}\hat{v}_r =0.
\label{104}
\end{equation}
%
The corresponding dispersion relation is therefore given by
\begin{equation}
\lambda=\pm 1,
\label{105}
\end{equation}
which represents, the stable epicyclical  oscillations in the disc plane, since for Keplerian rotation, the
epicyclical frequency equals to Keplerian one, $\bar{\chi}(\rho)= \bar{\Omega}(\rho)$. As $\rho$  and $\eta$
are mere parameters each ring  $\rho=const$, $\eta=const$  vibrates independently in its own plane. If
viscous stresses are taken into account, an axial profile is imposed due to the mutual shear stresses
between the rings and each entire cylindrical shell vibrates independently. This is indeed the case
that has been solved in [\cite{Umurhan et al. 2006}; \cite{Rebusco et al. 2009}); \cite{Shtemler et al. 2010}].
As the axial and radial coordinates play the role of passive parameters, the eigenfunctions of the
inertia-Coriolis modes are determined up to arbitrary amplitude $A(\rho, \eta)$  from
(\ref{103})-(\ref{105}):
\begin{equation}
\hat{v}_\theta=\mp i\frac{1}{2}\hat{v}_r =\mp i\frac{1}{2} A(\rho, \eta)
.
\label{106}
\end{equation}
Since any special form of the function   represents some given set of initial conditions, the following
self-similar separable form of the planar velocities is considered as an example:
\begin{equation}
A(\rho, \eta) = F(\rho) G(\eta),
\label{107}
\end{equation}
where the $F(\rho)$  and $G(\eta)$  are arbitrary functions.

{\it{Magnetosonic modes}}.
For frequencies that are different from the eigenvalues of inertial waves, namely
 $\lambda\equiv \omega/\Omega(\rho)=\pm 1$  in (\ref{105}), both the perturbed in-plane components of the velocity as well as the radial component of the perturbed magnetic field are zero, i.e.
 $\hat{v}_r=\hat{v}_\theta=\hat{b}_r\equiv0$. Therefore, the perturbed axial velocity, number density, as well as the toroidal magnetic field are described by the set of equations that governs the dynamics of the magnetosonic waves:
\begin{equation}
[1+\frac{1}{\bar{\beta}_\theta \bar{\nu}(\eta)}]
\frac{d^2 \hat{v}_z}{d \eta^2 }
-\eta  \frac{d\hat{v}_z}{d \eta }
+(\lambda^2-1) \hat{v}_z
=0,
\label{108}
\end{equation}
\begin{equation}
-i\lambda \frac{\hat{\nu}}{\bar{\nu}(\eta)}
+\frac{d \hat{v}_z}{d \eta }-\eta \hat{v}_z
=0,
\label{109}
\end{equation}
\begin{equation}
-i\lambda \hat{b}_\theta
+\frac{d \hat{v}_z}{d \eta}=
0,
\,\,\,
\label{110}
\end{equation}
is subject to the boundary conditions  at infinity of the least possible divergence for the axial velocity and the vanishing conditions for the in-plane magnetic-field components.
%
Partial solutions  are presented below .

{\it{(i) The pure hydrodynamic system}}.
First start with a simple exact solution for the pure hydrodynamic system. Accounting to Eq. (\ref{100}) this yields
%
\begin{equation}
\frac{d^2 \hat{v}_z}{d \eta^2 }
-\eta  \frac{d \hat{v}_z}{d \eta }
+(\lambda^2-1) \hat{v}_z
=0,
\label{112}
\end{equation}
\begin{equation}
-i\lambda \frac{\hat{\nu}}{\bar{\nu}(\rho)}
+\frac{d \hat{v}_z}{d \eta }-\eta \hat{v}_z=0,
\label{113}
\end{equation}
complemented by the  boundary condition  of  least possible divergence  at infinity for the axial velocity.
That problem is degenerated to the sound waves, and coincides with the case
of the poloidal-dominated equilibrium  magnetic field, and its solution is:
 \begin{equation}
\lambda=\pm \Lambda_S=\pm\sqrt{m+1},\,\,\,
\hat{v}_z=-i \lambda W(\rho) H_{m}(\eta),\,\,\,
\frac{\hat{\nu}}{\bar{\nu}}=  W(\rho) H_{m+1}(\eta)\,\,\,\,\,\,\,
\mbox{at}\,\,\,\,\eta=\pm\infty.
\label{114}
\end{equation}
Here $\lambda$ are the radius-independent  eigenvalues; $W(\rho)$ are the radius-dependent amplitude factors;   $H_m(\eta),\, m=0,1,2,...$ are the Hermite polynomials.

{\it{(ii) The limit of large  plasma beta in magnetized discs.}}
The WKB approximation is employed now. In general, it may be applied for a wide range of plasma beta values. However, in order to simplify calculations, we restrict ourselves here by its application to the limit of large but finite values of the  plasma beta. Such approach is necessary as a simple asymptotic expansion in large plasma beta $\bar{\beta}_\theta$ fails at large $\eta$  (small $\bar{\nu}(\eta)$) due to the non-uniform limit of $\bar{\beta}_\theta \bar{\nu}(\eta)$  in Eq. (\ref{108}).   As a first step the following  equation for  $\hat{b}_\theta$ is derived from the system (\ref{108}) - (\ref{110})
\begin{equation}
 \frac{d^2 \hat{b}_\theta}{d \eta^2 }
-\eta  \frac{\bar{\beta}^2_\theta \bar{\nu}(\eta)-1}{\bar{\beta}^2_\theta \bar{\nu}(\eta)+\bar{\beta}_\theta }\frac{d \hat{b}_\theta}{d \eta }
+(\lambda^2-1)
\frac{\bar{\beta}_\theta \bar{\nu}(\eta)}{1+\bar{\beta}_\theta \bar{\nu}(\eta)}
 \hat{b}_\theta
=0,
\label{115}
\end{equation}	
with the following boundary conditions:
\begin{equation}
 \hat{b}_\theta=0\,\,\,\,\,\,\,\,\,\,\,\,\,\,\,\,\mbox{at}\,\,\,\,\,\eta=\pm\infty.
\label{116}
\end{equation}
To reduce Eq.  (\ref{115}) to the form appropriate for WKB approximation, the dependent variable is transformed in the following way in order to eliminate terms with first order derivatives:
\begin{equation}
\hat{b}_\theta=Q \int \mu d\eta,\,\,\,\,\,
\mu = \frac{\eta }{2\bar{\beta}_\theta} \frac{\bar{\beta}^2_\theta \bar{\nu}(\eta)-1}{\bar{\beta}_\theta \bar{\nu}(\eta)+1 }.
\label{117}
\end{equation}
Then
\begin{equation}
 \frac{d^2 Q}{d \eta^2 }
+\varkappa^2(\eta) Q
=0,
\label{118}
\end{equation}	
where
\begin{equation}
\varkappa^2(\eta)=-\eta^2
\frac{1+\bar{\beta}_\theta  }{2\bar{\beta}_\theta }
 \frac{\bar{\beta}_\theta \bar{\nu}(\eta)}{[1+\bar{\beta}_\theta \bar{\nu}(\eta)]^2}
- \frac{1}{2\bar{\beta}_\theta }
 \frac{1-\bar{\beta}^2_\theta \bar{\nu}(\eta)}{1+\bar{\beta}_\theta \bar{\nu}(\eta)}
 -\eta^2
\frac{1}{4\bar{\beta}^2_\theta }
[\frac{1-\bar{\beta}^2_\theta \bar{\nu}(\eta)}{1+\bar{\beta}_\theta \bar{\nu}(\eta)}]^2
 +(\lambda^2-1)
\frac{\bar{\beta}_\theta \bar{\nu}(\eta)}{1+\bar{\beta}_\theta \bar{\nu}(\eta)}.
  \label{119}
\end{equation}
The approximate solutions appropriate to real and imagine  $\varkappa(\eta)$, at which the solution oscillates and decrease, respectively, are as follows (see e.g. \cite{Migdal 1977}):
\begin{equation}
Q=\frac{W}{\sqrt{\varkappa(\eta)}}\exp(\pm i\int \varkappa(\eta) d\eta\,\,\,\,\,\,\,\mbox{and}\,\,\,\,\,\,\,
Q=\frac{W}{\sqrt{\mid\varkappa(\eta)\mid}}\exp(\pm \int \mid\varkappa(\eta)\mid d\eta.
\label{120}
\end{equation}	
The regions of each of the solutions given in Eq. (\ref{120}) are determined by the turning points which are the zeros of  $\varkappa(\eta)$.
It can be easily seen that large-plasma beta approximation is uniformly valid in the region between the turning points, and violated at large  $\eta$ due to uncertainty in the term $\bar{\beta}_\theta \bar{\nu}(\eta)$  in (\ref{119}) for exponentially vanishing  $\bar{\nu}(\eta)=\exp(-\eta^2/2)$ at $\eta \to \pm \infty$  and  $\bar{\beta}_\theta \sim \beta\gg 1$. Since application of the WKB method is justified asymptotically at large frequency $\lambda$, Eq. (\ref{119}) is considered for large but fixed values of $\bar{\beta}_\theta $  and  $\lambda$,  and for such large  $\eta$ that both   $\bar{\beta}_\theta \bar{\nu}(\eta)$ and  $\bar{\beta}^2_\theta \bar{\nu}(\eta)$ are small. Then Eq. (\ref{119}) yields  to leading order
\begin{equation}
\varkappa^2(\eta)\approx -
\frac{1 }{2\bar{\beta}_\theta }(1+\frac{\eta^2 }{2\bar{\beta}_\theta }),\,\,\,\,\,
 \mu (\eta)\approx -
\frac{1}{2\bar{\beta}_\theta }\eta.
  \label{121}
\end{equation}
 Choosing the minus sign to satisfy the boundary condition (\ref{116})  in the approximate solution (\ref{120}) for $\mid\eta\mid>\mid\eta_*\mid$, yields  the following form  for sufficiently large  $\eta$:
\begin{equation}
\hat{b}_\theta\sim\frac{1}{\mid \varkappa \mid}\exp(-\frac{\eta^2}{4\bar{\beta}_\theta }
\pm \int_{\eta_*}^{\eta} \mid \varkappa \mid d\eta)=\frac{2\bar{\beta}_\theta }{\eta^2}\exp(-\frac{\eta^2}{2\bar{\beta}_\theta }).
\label{122}
\end{equation}
To further simplify the calculations, assume additionally that  $\bar{\beta}_\theta \gg 1$, then the large beta limit is uniformly valid within the interval  $\mid \eta \mid \leq \mid \eta_*\mid$, and   $\varkappa$ and   $\mu $  are given by
\begin{equation}
\varkappa^2(\eta)\approx -\frac{1}{4}(\eta^2-\eta^2_*),\,\,\,\,
 \mu (\eta)\approx \frac{1}{2}\eta,
  \label{123}
\end{equation}
where  $\eta_*=\pm \sqrt{4\lambda^2 -2}$ are the turning points of the equation (\ref{118}) which separate the regions of oscillatory and exponentially decreasing behavior of the solution. Thus,  $\varkappa=0$ at  $\mid \eta \mid = \mid \eta_*\mid$,  $\varkappa^2$ is positive within the inner region$\mid \eta \mid \leq \mid \eta_*\mid$    and negative otherwise within the outer region. To completely determine the problem solution, the conventional WKB approximation applies the matching conditions to the solutions (\ref{120}) in the inner ($\mid \eta \mid \leq \mid \eta_*\mid$) and outer ($\mid \eta \mid \geq \mid \eta_*\mid$) regions in order to obtain the coefficients in (\ref{120})  and the eigenvalue equation. The result is the following Bohr-Zommerfeld condition [\cite{Landau and Lifshits 1997}] that determines the dispersion relation
\begin{equation}
\int_{-\eta_*}^{+\eta_*} \varkappa  d\eta=\pi(m+\frac{1}{2}),
\label{124}
\end{equation}
where  $\varkappa$ is given by (\ref{123}), and $m$ is the number of zeros of the solution within the inner region. Substituting  (\ref{123}) into  (\ref{124}) yields
 \begin{equation}
\lambda =\pm \sqrt{m+1}.
\label{125}
\end{equation}
The eigenvalues
(\ref{125}) exactly coincide with those in  (\ref{114}) obtained with no restrictions on the value of axial wave numbers $m$  for the pure hydrodynamic system.

Finally note that although the WKB method is valid asymptotically for large frequencies, $\lambda\gg 1$ (i.e. large plasma beta or, alternatively, large axial wave numbers, $m$), it frequently yields a fair approximation up to finite values $\lambda\sim 1$ [see e.g. \cite{Migdal 1977}]. This allows expecting that eigenvalues (\ref{125}) will be appropriate even for low axial wave numbers $m=1,2,\dots$.

{\it{(iii) The limit of small  plasma beta  in magnetized discs.}}
Consider the problem (\ref{108})  -(\ref{110})  in the limit of small plasma beta $\bar{\beta}_\theta(\rho) \sim \bar{\beta}\ll 1$. Assuming simultaneously that the eigenvalue $\lambda$  is large, so that  $\lambda^2 \bar{\beta}_\theta\sim\bar{\beta}^0$, Eq. (\ref{108})   yields to leading order in small $\bar{\beta}$
\begin{equation}
\frac{d^2 \hat{v}_z}{d \eta^2 }
+\lambda^2 \bar{\beta}_\theta \bar{\nu} (\eta) \hat{v}_z=0.
\label{126}
\end{equation}
Replacing as in Section 3.3 the steady-state isothermal density distribution $\bar{\nu} (\eta) =\exp(-\eta^2/2)$  by the
hyperbolic function $\bar{\nu} (\eta)=\mbox{sech}^2(b\eta)$ with $b=\sqrt{2/\pi}$, and introducing the auxiliary variable
$\xi=\mbox{tanh} (b\eta)$, transform Eq. (\ref{126}) to the following form:
\begin{equation}
L \hat{v}_z+\bar{\lambda}^2 \hat{v}_z=0,\,\,\,\,
\label{127}
\end{equation}
where $L$  is the Legendre differential operator of second order,
$$
L\equiv \frac{d\,}{d \xi}[(1-\zeta^2) \frac{d\,}{d \xi}],\,\,\,
\,\,\,\,\,\,\,\,\,\,\,\,\,\,\,\,\,\,\,\,\,\,\,\,\,\,\,\,\,\,\,\,\,\,\,\,\,\,\,\,\,\,\,\,\,\,\,\,\,\,\,\,\,\,
\,\,\,\,\,\,\,\,\,\,\,\,\,\,\,\,\,\,\,\,\,\,\,\,\,\,\,\,\,\,\,\,\,\,\,\,\,\,\,\,\,\,\,\,\,\,\,\,\,\,\,\,\,\,
\,\,\,\,\,\,\,\,\,\,\,\,\,\,\,\,\,\,\,\,\,\,\,\,\,\,\,\,\,\,\,\,\,\,\,\,\,\,\,\,\,\,\,\,\,\,\,\,\,\,\,\,\,\,
\,\,\,\,\,\,\,\,\,\,\,\,\,\,\,\,\,\,\,\,\,\,\,\,\,\,\,\,\,\,\,\,\,\,\,\,\,\,\,\,\,\,\,\,\,\,\,\,\,\,\,\,\,\,
\,\,\,\,\,\,\,\,\,\,\,\,\,\,\,\,\,\,\,\,\,\,\,\,\,\,\,\,\,\,\,\,\,\,\,\,\,\,\,\,\,\,\,\,\,\,\,\,\,\,\,\,\,\,
\,\,\,\,\,\,\,\,\,\,\,\,\,\,\,\,\,\,\,\,\,\,\,\,\,\,\,\,\,\,\,\,\,\,\,\,\,\,\,\,\,\,\,\,\,\,\,\,\,\,\,\,\,\,
\,\,\,\,\,\,\,\,\,\,\,\,\,\,\,\,\,\,\,\,\,\,\,\,\,\,\,\,\,\,\,\,\,\,\,\,\,\,\,\,\,\,\,\,\,\,\,\,\,\,\,\,\,\,
\,\,\,\,\,\,\,\,\,\,\,\,\,\,\,\,\,\,\,\,\,\,\,\,\,\,\,\,\,\,\,\,\,\,\,\,\,\,\,\,\,\,\,\,\,\,\,\,\,\,\,\,\,\,
$$
and $\bar{\lambda}^2$  is given by:
\begin{equation}
\bar{\lambda}^2 = \frac{1}{b^2} \lambda^2(\rho) \bar{\beta}_\theta(\rho)\sim \bar{\beta}^0.
\label{128}
\end{equation}
Imposing now that $\hat{v}_z$ diverges polynomially at most when $\eta\to \infty$, leads to the conclusion that the solution of Eq.
(\ref{128}) for $\hat{v}_z$  is proportional to the Legendre polynomials $P_m(\xi)$:
\begin{equation}
\frac{\hat{v}_z}{W(\rho)}= P_m(\xi)\sim \bar{\beta}^{0},\,\,\,\, \bar{\lambda}^2 =m(m+1)\sim \bar{\beta}^{0},\,\,\,\,m=1,2,....
\label{129}
\end{equation}
Equations (\ref{106}) - (\ref{107})
yield that $\hat{b}_\theta$  and $\hat{\nu}$  are of higher order in small plasma beta:
$$
\frac{\hat{b}_\theta}{W(\rho)} =-i\sqrt{\frac{m}{m+1}}
 [P_{m-1}(\xi)-\xi P_m(\xi)]\bar{\beta}_\theta^{1/2}\sim \bar{\beta}^{1/2},\,\,\,
 \,\,\,\,\,\,\,\,\,\,\,\,\,\,\,\,\,\,\,\,\,\,\,\,\,\,\,\,\,\,\,\,\,\,\,\,\,\,\,\,\,\,\,\,\,\,\,\,\,\,\,\,\,\,
\,\,\,\,\,\,\,\,\,\,\,\,\,\,\,\,\,\,\,\,\,\,\,\,\,\,\,\,\,\,\,\,\,\,\,\,\,\,\,\,\,\,\,\,\,\,\,\,\,\,\,\,\,\,
\,\,\,\,\,\,\,\,\,\,\,\,\,\,\,\,\,\,\,\,\,\,\,\,\,\,\,\,\,\,\,\,\,\,\,\,\,\,\,\,\,\,\,\,\,\,\,\,\,\,\,\,\,\,
\,\,\,\,\,\,\,\,\,\,\,\,\,\,\,\,\,\,\,\,\,\,\,\,\,\,\,\,\,\,\,\,\,\,\,\,\,\,\,\,\,\,\,\,\,\,\,\,\,\,\,\,\,\,
\,\,\,\,\,\,\,\,\,\,\,\,\,\,\,\,\,\,\,\,\,\,\,\,\,\,\,\,\,\,\,\,\,\,\,\,\,\,\,\,\,\,\,\,\,\,\,\,\,\,\,\,\,\,
\,\,\,\,\,\,\,\,\,\,\,\,\,\,\,\,\,\,\,\,\,\,\,\,\,\,\,\,\,\,\,\,\,\,\,\,\,\,\,\,\,\,\,\,\,\,\,\,\,\,\,\,\,\,
\,\,\,\,\,\,\,\,\,\,\,\,\,\,\,\,\,\,\,\,\,\,\,\,\,\,\,\,\,\,\,\,\,\,\,\,\,\,\,\,\,\,\,\,\,\,\,\,\,\,\,\,\,\,
$$
\begin{equation}
\frac{\hat{\nu}}{W(\rho)} =\frac{i}{b}\sqrt{\frac{m}{m+1}}(1-\xi^2)
 [(\xi+\frac{1}{mb}\mbox{arctanh}\xi)P_{m}(\xi)- P_{m-1}(\xi)]\bar{\beta}_\theta^{1/2}\sim \bar{\beta}^{1/2}.
\label{130}
\end{equation}
Since $P_{m}(1)=1$  and  $P_{m}(-1)=(-1)^m$  for all $m$,  $\hat{b}_\theta$ satisfies the vanishing boundary condition at infinity $\eta\to\pm\infty$  ($\xi\to\pm 1$).

Thus, summarizing, the main conclusion from both limits (high and low plasma beta) is that the inertia-Coriolis waves are decoupled from the Alfv\'en waves, which henceforth leads to the complete stabilization of the MRI's. This, as will be subsequently seen, leaves the stage exclusively to the non modal algebraic growth mechanism.  Remarkably that both limits assume sufficiently large values of the eigenvalues (scaled frequensies $\lambda$): in the limit of large $\beta$  in (\ref{125}) due to high frequency assumption inherent to WKB, while in opposite limit of small  $\beta$ in (\ref{129}), because to asymptotic behavior of $\lambda\sim \beta^{-1/2}$.

\subsection{Non-resonantly and resonantly driven magnetosonic modes}
There are two mechanisms responsible for algebraic time-growth of perturbations, namely non-resonant and resonant excitation of vertical magnetosonic waves by planar inertia-Coriolis modes. This fact for the pure hydrodynamic adiabatic discs was established in [\cite{Umurhan et al. 2006}; \cite{Rebusco et al. 2009}); \cite{Shtemler et al. 2010}].  As shown below, both mechanisms are also relevant in a modified form for  vertically-isothermal  magnetized  discs. Due to the rotation shear effect and the presence of radial derivatives in the system (\ref{94}) -(\ref{100}), the amplitude of the magnetosonic modes driven by the inertia-Coriolis modes grows linearly in time in the case when the magnetosonic mode frequency is not equals to the inertial-Coriolis one. While for the same frequencies of the magnetosonic and inertial-Coriolis modes, the latter grows quadratically in time.

To demonstrate linear and quadratic in time growth of the perturbations, both non-resonantly and resonantly driven modes are illustrated below  by simple explicit solutions. The non-resonantly driven magnetosonic waves are considered in the limit of small toroidal plasma beta and for pure hydrodynamic system, as typical examples. Noting that the resonantly driven magnetosonic modes may exist for arbitrary  frequencies, we restrict ourselves by considering a pure hydrodynamic system which is characterized by simple explicit solutions with no limitation on the frequency value.
%


%
The  driving modes are characterized by Eqs. (\ref{102}) and (\ref{106}). As a result, the driven  magnetosonic or sound modes in magnetized or magnetic-field free  discs, i.e. (\ref{98}) or (\ref{100}), respectively,  are described by the following expressions:
\begin{equation}
\{v_r,\,\,v_\theta,\,\,\,b_\theta\}=\exp[-i\lambda\bar{\Omega}(\rho)t]
\{\hat{v}_r,\,\,\hat{v}_\theta,\,\,\,\hat{b}_\theta\}(\rho,\eta,t).
\label{131}
\end{equation}
Inserting then Eq. (\ref{131}) into the system of equations ((\ref{93})-( (\ref{100}) and keeping only terms that are proportional to $t$  on their right hand sides yield
\begin{equation}
%
\frac{1}{\bar{\Omega}^2}  \frac{\partial^2 v_z}{\partial t^2 }
-(1+\frac{1}{\bar{\beta}_\theta \bar{\nu}(\eta)})\frac{\partial^2 v_z }{\partial \eta^2 }
+\eta\frac{\partial v_z }{\partial \eta } + v_z=\mp i t\exp[-i\lambda\bar{\Omega}(\rho)t](1+\frac{1}{\bar{\beta}_\theta \bar{\nu}(\eta)})
\frac{d\bar{\Omega}}{d\rho}\frac{\partial\hat{v}_r}{\partial \eta}
,
\label{132}
\end{equation}
\begin{equation}
\frac{1}{\bar{\Omega}}  \frac{\partial }{\partial t}[\frac{\nu}{\bar{\nu}(\eta)}]
+\frac{\partial v_z}{\partial \eta }-\eta v_z
=\pm i t\exp[\mp i\bar{\Omega}(\rho)t]\,
\frac{d\bar{\Omega}}{d\rho} \hat{v}_r,
\label{133}
\end{equation}
\begin{equation}
\frac{1}{\bar{\Omega}}   \frac{\partial b_\theta}{\partial t}
+\frac{\partial v_z}{\partial \eta}=
\pm i t\exp[\mp i\bar{\Omega}(\rho)t]
\frac{d \bar{\Omega}}{d
\rho}\hat{v}_r.
\label{134}
\end{equation}
The  problems (\ref{132})-(\ref{134}) is subject to the  boundary condition of  least possible divergence  for the axial velocity, and for the magnetized discs  complemented by
 the vanishing conditions of the toroidal magnetic field at $\eta=\pm\infty$.
  As is indeed seen from that system, the magnetosonic wave is driven by the inertial wave, while the linear growth in time is entirely due to effect of the rotation shear, and disappears for  $d\bar{\Omega}/d\rho=0$.

As was mentioned above  the two limits of small toroidal plasma beta as well as  pure hydrodynamic system are considered analytically. If the scaled eigen-frequency of the inertia-coriolis waves (\ref{109}),  $\lambda=\pm 1$, do not coincide with the eigen-frequencies of the magnetosonic waves (\ref{119}),  $\lambda=\pm\sqrt{m+1}$,  i.e. at  $m\neq 0$, non-resonantly driven magnetosonic modes are excited. The stable magnetosonic modes may be excited resonantly by the inertial modes if any pair of respective eigen-values (\ref{109}) with $\lambda=\pm 1$  and (\ref{119}) with  $\lambda=\pm\sqrt{m+1}$, coincide. It is easy to see that this may happen only for  $m=0$.

{\it{(i) Non-resonantly driven  magnetosonic modes in  the  limit of small  plasma beta}}.
At vanishing plasma beta, $\beta$,   in the leading order approximation Eq. (\ref{132})  leads to a degenerate quasi-steady problem for $\hat{v}_z$  on the Keplerian time scale. Twice integrating Eq. (\ref{132}) in   $\eta$ and setting arbitrary integration constants to zero, yield:

\begin{equation}
\hat{v}_z=\pm i t\frac{d \bar{\Omega}}{d
\rho}\int_0^\eta \hat{v}_r d\eta .
\label{135}
\end{equation}
The rest unknown number density and  toroidal magnetic field  can be found at known $\hat{v}_z$  from non-degenerate unsteady relations (\ref{133}) -(\ref{134}).


{\it{(ii) Non-resonantly and resonantly driven magnetosonic modes in pure hydrodynamic discs}}.
Setting $\bar{\beta}_\theta\equiv\infty$,  $\hat{v}_r$  on the right-hand sides of (\ref{132}) -(\ref{134})
may be expanded in terms of the complete set of the
 spatial eigen-functions for the magnetosonic mode, namely, in the Hermite polynomials:
\begin{equation}
\hat{v}_r(\rho,\eta)=\sum_{m=0}W_{r,m}(\rho)H_m(\eta).
\label{136}
\end{equation}
Examining the effect of a single term in the expansion above and utilizing the following recursion relation for the Hermite polynomials:
$$
\frac{d H_m }{d \eta }=m H_{m-1}(\eta),\,\,\,\,\,\,\,\,\,\,m=1,2,\dots,
\,\,\,\,\,\,\,\,\,\,\,\,\,\,\,\,\,\,\,\,\,\,\,\,\,\,\,\,\,\,\,\,\,\,\,\,\,\,\,\,\,\,\,\,\,\,\,\,\,
\,\,\,\,\,\,\,\,\,\,\,\,\,\,\,\,\,\,\,\,\,\,\,\,\,\,\,\,\,\,
\,\,\,\,\,\,\,\,\,\,\,\,\,\,\,\,\,\,\,\,\,\,\,\,\,\,\,\,\,\,\,\,\,\,\,\,\,\,\,\,\,\,\,\,\,\,\,\,\,\,\,\,\,\,\,\,\,\,\,\,
\,\,\,\,\,\,\,\,\,\,\,\,\,\,\,\,\,\,\,\,\,\,\,\,\,\,\,\,\,\,
\,\,\,\,\,\,\,\,\,\,\,\,\,\,\,\,\,\,\,\,\,\,\,\,\,\,\,\,\,\,\,\,\,\,\,\,\,\,\,\,\,\,\,\,\,\,\,\,\,\,\,\,\,\,\,\,\,\,\,\,
\,\,\,\,\,\,\,\,\,\,\,\,\,\,\,\,\,\,\,\,\,\,\,\,\,\,\,\,\,\,
\,\,\,\,\,\,\,\,\,\,\,\,\,\,\,\,\,\,\,\,\,\,\,\,\,\,\,\,\,\,
$$
substituting the exponential anzatz (\ref{131}) into Eq. (\ref{100}), and keeping only terms that are proportional to $t$  on its right hand side yield
%
\begin{equation}
%
\frac{1}{\bar{\Omega}^2}  \frac{\partial^2 v_{z,m-1}}{\partial t^2 }
-\frac{\partial^2 v_{z,m-1} }{\partial \eta^2 }
+\eta\frac{\partial v_{z,m-1} }{\partial \eta } + v_{z,m-1}=\mp i t\exp[\mp i\bar{\Omega}(\rho)t]
\frac{d\bar{\Omega}}{d\rho} W_{r,m-1}(\rho) m H_{m-1}(\eta),
\label{137}
\end{equation}
where for $m=1$  the right hand side of Eq. (\ref{137}) is a solution of the homogeneous part of the equation and hence represents a resonant driving force.  For any other value of $m$ the magnetosonic waves are driven non-resonantly.
The typical non-resonant ( $m=2$) and resonant ($m=1$ )
 solution of Eq. (\ref{137}) may be written in the following form:
\begin{equation}
v_{z,m-1}=\hat{v}_{z,m-1}(\rho,\eta,t)\exp[\mp i\bar{\Omega}(\rho)\theta]=\mp i \exp[\mp i\bar{\Omega}(\rho)t]W_{z,m-1}(\rho,t)
 H_{m-1}(\eta),
\label{138}
\end{equation}
\begin{equation}
 W_{z,1}(\rho,t)=(i t \bar{\Omega})W_{z,1}^{(1)}(\rho)+W_{z,1}^{(0)}(\rho)\ \  \ \ \ \   \  \ \ \ \
\ \ \ \ \ \ \ \ \ \ \ \ \  \    \mbox{for}\ \ \ \ \  m=2, \ \  \ \ \ \
\label{138a}
\end{equation}
\begin{equation}
 W_{z,0}(\rho,t)=\frac{ (i t \bar{\Omega})^2}{2}W_{z,0}^{(2)}(\rho)+(i t \bar{\Omega})W_{z,0}^{(1)}(\rho)
\ \ \ \ \ \ \ \ \ \ \ \ \ \ \ \ \mbox{for}\ \ \ \ \  \    m= 1. \ \
\label{138b}
\end{equation}
Here
 the coefficients are determined through an arbitrary amplitude of the radial velocity $W_{r,1}(\rho)$
 for non-resonant ($ m= 2$) and resonant ($ m= 1$) system, respectively, are as follows
\begin{equation}
\frac{ W_{z,1}^{(1)}(\rho)}{W_{r,1}(\rho)}= - i \frac{2}{\bar{\Omega}} \frac{d\bar{\Omega})}{d\rho}
,\,\,\,\,\,\,
\frac{ W_{z,1}^{(0)}(\rho)}{W_{r,1}(\rho)}=\pm i \frac{2}{\bar{\Omega}} \frac{d\bar{\Omega}}{d\rho}
\ \  \ \ \ \   \  \ \ \ \
\ \ \ \ \ \ \      \mbox{for}\ \ \ \ \  m=2, \ \  \ \ \ \
\label{139}
\end{equation}
\begin{equation}
\frac{ W_{z,0}^{(2)}(\rho)}{W_{r,1}(\rho)}=\mp i \frac{1}{\bar{\Omega}} \frac{d\bar{\Omega})}{d\rho}
,\,\,\,\,\,\,
\frac{ W_{z,0}^{(1)}(\rho)}{W_{r,1}(\rho)}=- i\frac{1}{2} \frac{1}{\bar{\Omega}} \frac{d\bar{\Omega}}{d\rho}
\ \ \ \ \ \ \ \ \ \ \ \ \ \ \ \ \mbox{for}\ \ \ \ \  \    m= 1. \ \
\label{140}
\end{equation}

The corresponding expressions for number density and  toroidal magnetic field     may be derived from the rest equations of the system (\ref{133}) -(\ref{134}).

\section{Summary and discussion}
A comprehensive asymptotic analysis in small aspect ratio of the disc,  $\epsilon$, has been carried out of the response of  thin vertically-isothermal Keplerian discs to small magnetohydrodynamic perturbations. Two regimes of axisymmetric instability  have been identified depending on the type of equilibria in small aspect ratio of the disc,  $\epsilon$. The first is developed in the axially dominated equilibrium magnetic configurations,  $\bar{B}_z\sim\epsilon^0$, and is excited at a relatively low level of hydrodynamic perturbations in the disc plane  $V_r',\,V_\theta'\sim\epsilon^1$, while the second regime occurs in the toroidally dominated magnetic fields,  $\bar{B}_\theta\sim\epsilon^0$, at relatively high level of hydrodynamic perturbations, $V_r',\,V_\theta'\sim\epsilon^0$  (the axial velocity perturbations are in both regimes of the same orders,   $V_z'\sim\epsilon^1$,
see Tables 1 and 2). As a result of that scaling, the pure hydrodynamic limit is achievable only within the second regime.  Indeed, it is demonstrated that as distinct from the first regime, the second regime can not produce spectrally unstable modes, but excites a weak algebraic non-modal growth in time. The inertia-Coriolis driven magnetosonic mode leads to their non-resonant and resonant coupling that induces, respectively, the linear and quadratic in time growth of perturbations.

It should be emphasized though that that result is restricted to axisymmetric perturbations and nonaxisymmetric ones may give rise to spectral instabilities in the dominant toroidal field case as well
 as indeed is the case for magnetized Taylor-Couette flows  (see Rudiger et al. 2007).
Indeed, \cite{Terquem and Papaloizou 1996} have derived sufficient conditions for such instabilities in some sense of a thin disk limit. If such instabilities indeed exist they are of a transient nature as the radial wave number grows and drives the system out of the instability regime, \cite{Balbus and Hawley 1992}. In addition, as demonstrated in Sections 3 and 4 spectrally unstable   perturbations may occur if the Lorentz force is of the same order as the inertia terms. This means that the spectrally unstable  non-axisymmetric perturbations are expected to be of higher order in   $\epsilon$ than those considered in Section 4, that grow linearly and quadratically with time.  Thus, it takes full nonlinear numerical calculations to determine which perturbations are more significant, namely, non-axisymmetric ones that grow exponentially from a very low level with a decreasing growth rate, or linearly and quadratically growing symmetric perturbations that grow from initial disturbances of much higher amplitudes.

For perturbations of type I the main accent is made on the dominant poloidal magnetic field, while 
the general case of comparable poloidal and toroidal components of the equilibrium magnetic field is discussed in Appendix A.
Explicit solutions of the stability problem for the dominant equilibrium poloidal magnetic field  of type I are obtained by replacing the true isothermal density vertical steady-state distribution $\bar{\nu}(\eta)=\exp(-\eta^2/2)$  by the hyperbolic function  $\bar{\nu}(\eta)=\mbox{sech}^2(b\eta)$  which has the similar shape and the same total mass of the disc. This model profile represents some true equilibrium that is obtained from a slightly different gravitational potential. The model eigenfunctions are exact solutions of the model very close to those obtained approximately by WKB from the true problem [\cite{Liverts and Mond 2009}], but they form the full family of explicit, simple and orthogonal eigenfunctions of the model problem. Such properties are significant for the consequent study of non-linear development of the instability.

A study of the stability of isothermal disks has been presented before in \cite{Gammie and Balbus 1994} and \cite{Latter et al. 2010}, who have also utilized the model density profile  $\mbox{sech}^2\eta$.  The main differences between those two works and the current investigation are:

(i)	\cite{Gammie and Balbus 1994} and \cite{Latter et al. 2010} have approached the stability problem of the disk within the local setting of a shearing box described by local Cartesian coordinates. In contrast, in the current work the entire disk is considered and is described by the natural cylindrical coordinates with the aid of asymptotic analysis in the natural small parameter of the problem, namely, the ratio of the thickness of the disk to its characteristic radius $\sim\epsilon$.

(ii)	Stemming from the shearing box local approach, the solutions for the perturbations that are given in  	 \cite{Gammie and Balbus 1994} and \cite{Latter et al. 2010}  are independent of the radial-like coordinate. In the current work, however, the radial derivatives drop out of the equations due to the asymptotic analysis in small  $\epsilon$. Consequently, the radial coordinate is a mere parameter in the subsequent calculations and conclusions. Thus, the result is that perturbations are excited on each separate cylindrical shell, independently of the other shells, with amplitude that depends arbitrarily on the radius through some initial conditions. As a result, it comes out rigorously that the stability conditions depend on the radius, and different rings within the disk have different stability properties, like number of unstable modes as well as growth rates. 	

(iii)	The current work includes also analysis of the acoustic spectrum.  Though stable in the case of purely axial magnetic field (explicit expressions are derived for the eigenvalues as well as for the eigenfunctions), as is sketched in Appendix A for large plasma beta, a toroidal component of the same order of magnitude as the axial one leads to the driving of acoustic modes by unstable MRI's.

(iv)		Figure 1 demonstrates that best choice of the parameter  $b$ is obtained from the requirement of identical 
total mass of the disc
for the two density profiles,    $\bar{\nu}(\eta)=\exp(-\eta^2/2)$  and $\bar{\nu}(\eta)=\mbox{sech}^2(b\eta)$.

In addition a qualitative analysis of the  influence of the equilibrium toroidal magnetic field of type I on the disc stability is carried out in Appendix A in the limit of large plasma beta on two characteristic scales of AC's  and  MS's  modes.
In that case as distinct from the small plasma beta ($\beta\lesssim  1$), the AC and MS modes are decoupled.
On the characteristic scale of AC's mode, this leaves the resulting dispersion relation the same as in the case of the dominant poloidal magnetic field (zero toroidal magnetic field), and the influence of the toroidal magnetic field is reduced to excitation of the AC-driven MS mode. On the characteristic scale of MS's mode, a stable MS mode decouples from the AC mode and the influence of the toroidal magnetic field is reduced to excitation of the stable MS-driven AC mode.

In the present work the  model of the non-modal instability [\cite{Shtemler et al. 2010}] for pure hydrodynamic adiabatic discs is adopted  for more realistic vertically-isothermal discs with diffused density vanishing at infinity. The model is also extended on magnetized discs with  type II equilibria. Within that model the eigenvalue problem has been solved in the characteristic cases of small and large plasma beta as well as for pure hydrodynamic systems. Furthermore, the magnetosonic waves driven non-resonantly with inertia-Coriolis ones are considered in the limit of small toroidal plasma beta, as a typical example. Since in the limits of large and small plasma beta, the magnetosonic modes are excited with high scaled frequency,  $\lambda\gg 1$, they can not be resonantly driven by the low-frequency inertial-Coriolis mode with  $\lambda\pm 1$. By this reason the resonantly driven magnetosonic modes are calculated for only the pure hydrodynamic system, in order to illustrate a quadratic in time growth of perturbations. The rotation shear is found to be responsible for the linear and quadratic growth in time. Furthermore, compressible inertia-coriolis stable oscillations continuously pump energy from the sheared steady state equilibrium and transfer it (resonantly as well as non resonantly) to the continuously algebraically growing magnetosonic waves. Compressibility is indeed inevitable due to the supersonic rotation of the steady-state disc combined with its small vertical dimensions which make the vertical magnetosonic crossing time of the order of a rotation period. As distinct from MRI the non modal growth is exhibited for all admissible parameters of the system.

Note that the present study has been restricted by consideration of the axisymmetric perturbations only. In particular, for type II perturbations the adopted analysis is considered as the necessary step in achieving a complete picture of transition to turbulence in accretion discs. First, it provides the instability for the case of very weak magnetic fields close to pure hydrodynamic perturbations in the discs. Furthermore, since the magnetized discs are stable with respect to axisymmetric normal modes, an alternative scenario of an algebraic instability has been investigated. The present analysis of both resonantly and non-resonantly driven magnetosonic modes is interesting in the context of the well known scenario for sub-critical transition to turbulence \cite{Schmidt and Henningson 2001}, since the non-spectral mechanism of the power growth in time provides instability even for the system parameters of the discs stable with respect to normal modes. The considered scenario leaves to be valid for sub-critical parameters of the discs within more general model of non-axisymmetric MRI [e.g. \cite{Balbus and Hawley 1992}, \cite{Terquem and Papaloizou 1996}, \cite{Foglizzo and Tagger 1995}].


\appendix
\section{Stability for comparable in-plane equilibrium magnetic fields}\label{app A}

The general relations for comparable in-plane components of the equilibrium magnetic field of type I are presented below. Substituting the exponential anzatz (\ref{43})-(\ref{44}) into the general system (\ref{37})-(\ref{42}) results in the following system of linear ordinary differential equations (parametrically depending on the radial variable $r$) for the perturbed toroidal magnetic field and axial velocity, $\hat{b}_\theta$  and $\hat{v}_z  $:
\begin{equation}
L_{AC}\hat{b}_\theta=\frac{1}{2}\bar{S}(r)\bar{\beta}_z(r)\big{\{}
\frac{d}{d\eta} \left[\frac{1}{\bar{\nu}(\eta)}
\frac{d^2\hat{v}_z }{d\eta^2} \right]
-3(\lambda^2-1)\bar{\beta}_z(r)
\frac{d\hat{v}_z }{d\eta}  \big{\}},
\label{A1}
\end{equation}
\begin{equation}
L_{MS}\hat{v}_z =-i\lambda\frac{\bar{S}(r)}{\bar{\beta}_z(r)}
\frac{d}{d\eta} \left[\frac{\hat{b}_\theta}{\bar{\nu}(\eta)}
\right].
\label{A2}
\end{equation}
Here $\bar{S}(r)$  is the coupling coefficient between AC and MS modes,
\begin{equation}
L_{AC}\hat{b}_\theta\equiv\frac{d}{d\eta} \big{\{}\frac{1}{\bar{\nu}(\eta)}
\frac{d^2}{d\eta^2} \left[\frac{1}{\bar{\nu}(\eta)}\frac{d\hat{b}_{\theta}}{d\eta}\right]\big{\}}
+(3+2\lambda^2)\bar{\beta}_z(r)
\frac{d}{d\eta}  \left[\frac{1}{\bar{\nu}(\eta)}\frac{d\hat{b}_{\theta}}{d\eta}\right]
+ \lambda^2( \lambda^2-1)\bar{\beta}_z^2(r) \hat{b}_{\theta},
\label{A3}
\end{equation}
 \begin{equation}
L_{MS}\hat{v}_z \equiv
\frac{d^2 \hat{v}_z}{d \eta^2 }+\frac{1}{\bar{\nu}(\eta)}\frac{d \bar{\nu}(\eta)}{d \eta }\frac{d \hat{v}_z}{d \eta }
+\big{\{}\lambda^2+\frac{d }{d \eta }\big[\frac{1}{\bar{\nu}(\eta)}\frac{d \bar{\nu}(\eta)}{d \eta }
\big]\big{\}}\hat{v}_z
\equiv
\frac{d^2 \hat{v}_z}{d \eta^2 }-\eta\frac{d \hat{v}_z}{d \eta }+(\lambda^2-1)\hat{v}_z.
\label{A4}
\end{equation}
Equations (\ref{A1}) - (\ref{A2}) are complemented by the boundary conditions of the toroidal magnetic field vanishing and of the least possible divergence for the axial velocity at infinity. The operators   $L_{AC}$ and $L_{MS}$  are such that the two principal, AC and MS, modes satisfy the following homogeneous problems for zero equilibrium toroidal magnetic field (i.e. for zero coupling coefficient $\bar{S}(r)\equiv 0$)  with the same boundary conditions:
 \begin{equation}
L_{AC}\hat{b}_\theta=0,
\label{A5}
\end{equation}
 \begin{equation}
L_{MS}\hat{v}_z =0.
\label{A6}
\end{equation}
Below that  problem will be considered for large plasma beta on two characteristic scales of AC's  and  MS's  modes.

\subsection{Zero toroidal magnetic field. }
First consider the problem for zero coupling coefficient,  $\bar{S}(r)$. Note that if set  $\bar{S}(r)=0$, the equilibrium toroidal magnetic field has no input in the thin disc approximation under consideration, the problems for AC and MS modes are decoupled,  and  can be considered separately (as in Section 3 above).
The method of two-scale asymptotic expansions in large plasma-beta will be applied below, which is convenient for the qualitative analysis of the disc stability. The method is equivalent to WKB approach to the problem (\ref{A5}) for MRI mode that also implicitly assumes  $\beta>>1$. In that connection note that as is demonstrated in \cite{Liverts and Mond 2009} the WKB approximation for MRI mode well describes the exact numerical solution up to plasma beta  $\beta\sim1$ (see also Section 3 where the WKB approximation is mentioned very close to the exact explicit solution for model density profile). 

According to the method of two-scale asymptotic expansions [see e.g. \cite{Nayfeh 1973}], let us introduce the slow, $\bar{\eta}$ , and fast,   $\tilde{\eta}$, variables at  $\beta>>1$ for AC mode:
 \begin{equation}
\bar{\eta}=\eta,\,\,\,
\tilde{\eta}=\beta^a \int_0^\eta g(\eta)\eta,\,\,\,\,\,(a>0),\,\,\,\tilde
{\beta}_z(r)=\frac{\bar{\beta}_z(r)}{\beta}\sim\beta^0,\,\,\,\frac{d}{d\eta} =\beta^a \big{[}g(\bar{\eta})\frac{\partial}{\partial\tilde{\eta}}+\beta^{-a }\frac{\partial}{\partial\bar{\eta}} \big{]},\,\,\,\,
\label{A7}
\end{equation}
\begin{equation}
\hat{b}_\theta(\eta)=\hat{b}_{\theta,0}(\bar{\eta},\tilde{\eta})+
\beta^{-a}\hat{b}_{\theta,1}(\bar{\eta},\tilde{\eta})+\dots,\,\,\,
\lambda=\lambda_0+\beta^{-a }\lambda_1+\dots.
\label{A8}
\end{equation}
Applying  the principle of the least possible degeneration of the problem yields:
\begin{equation}
a=1/2.									
\label{A9}
\end{equation}
 To make the coefficients of the leading order equation (\ref{A1}) independent of the slow variable, set
\begin{equation}
g(\bar{\eta})=\bar{\nu}^{1/2}(\bar{\eta}).									
\label{A10}
\end{equation}
This yields
\begin{equation}
\frac{\partial^4\hat{b}_{\theta,0}}{\partial\tilde{\eta}^4}
+(3+2\lambda_0^2)\tilde{\beta}_z(r)\frac{\partial^2\hat{b}_{\theta,0}}{\partial\tilde{\eta}^2}
+\lambda_0^2( \lambda_0^2-1)\tilde{\beta}_z^2(r)  \hat{b}_{\theta,0}=0.
\label{A11}
\end{equation}
The vanishing at infinity solution of the linear equation (\ref{A11}) can be easily found explicitly to leading order in $\beta$ up to amplitude factors that depend on slow variable,  $\bar{\eta}$. The equation (\ref{A11}) evidently conserves all the principle derivatives by fast variable in the exact Eq. (\ref{A5}).  The dependence of the amplitude factors on the slow variable  are determined by the problem of the next order approximation in large plasma beta. To leading order in $\beta$, these two first leading order problems completed by the corresponding boundary conditions fully determine the approximate solution of the problem including the dispersion relation for the eigenvalue  $\lambda_0$. 

 %

\subsection{Comparable toroidal and poloidal magnetic fields at the characteristic scale of AC mode. }
 Let us now apply  the two-scale asymptotic expansions (\ref{A7}) - (\ref{A8}) in $\beta>>1$ to complete relations (\ref{A1}) - (\ref{A4}) for the mixed toroidal and poloidal magnetic fields on the characteristic scale of AC's mode. It is assumed as above in the case of separated AC mode that $a>0$, and additionally
\begin{equation}
\bar{S}(r)=\sqrt{\frac{\bar{\beta}_z(r)}{\bar{\beta}_\theta(r)}}\sim\beta^0,\,\,\,
 \hat{v}_z(\eta)=\beta^c f(\bar{\eta})[\hat{v}_{z,0}(\bar{\eta},\tilde{\eta})+
\beta^{-a}\hat{v}_{z,1}(\bar{\eta},\tilde{\eta})+\dots],\,\,\,
\lambda=\lambda_0+\beta^{-1 }\lambda_1+\dots.
\label{A12}
\end{equation}
(i) {\it{Leading  order approximation.} } The principle of the least possible degeneration of the problem yields:
\begin{equation}
a=1/2,\,\,\,c=-2.									
\label{A13}
\end{equation}
To make the coefficients of the resulting relations to leading order independent of the slow variable,  set
\begin{equation}
g(\bar{\eta})=\bar{\nu}^{1/2}(\bar{\eta}),\,\,\, f(\bar{\eta})=\bar{\nu}^{-3/2}(\bar{\eta}).									 \label{A14}
\end{equation}
This leaves Eq. (\ref{11}) valid for the AC mode
\begin{equation}
\frac{\partial^4\hat{b}_{\theta,0}}{\partial\tilde{\eta}^4}
+(3+2\lambda_0^2)\tilde{\beta}_z(r)\frac{\partial^2\hat{b}_{\theta,0}}{\partial\tilde{\eta}^2}
+ \lambda_0^2( \lambda_0^2-1)\tilde{\beta}_z^2(r) \hat{b}_{\theta,0}=0,
\label{A15}
\end{equation}
 while the MS mode is the AC-driven mode
\begin{equation}
\frac{\partial \hat{v}_{z,0}}{\partial \tilde{\eta}}=-i\lambda_0\frac{\bar{S}(r)}{\tilde{\beta}_z(r)}\hat{b}_{\theta,0}.									 \label{A16}
\end{equation}
Thus, in the limit of large plasma beta, the AC mode decouples on its characteristic scale from  the  MS mode, and the influence of the toroidal magnetic field leaves the problem for the AC mode the same as in the case of zero equilibrium toroidal field, and simultaneously  induces  the AC-driven MS mode.

(ii) {\it{On the next  order approximations.} } Note that the leading-order governing relation for $\hat{b}_{\theta,0}$ is of the order of  $\beta^2$, while the next-order  relation for   $\hat{b}_{\theta,1}$    is of the order of    $\beta^{3/2}$. Estimating the terms, which are proportional to the coupling coefficient  $\bar{S}(r)$  and induced by the equilibrium toroidal magnetic field,  note that they are of the order of    $\beta^{1/2}$, i.e. much less than    $\beta^{3/2}$. Hence the toroidal magnetic field has no input neither to the leading- nor to the next-order approximations, which  determine the resulting approximate solution of the problem. In particular, this leaves the resulting dispersion relation the same as in the case of zero equilibrium toroidal magnetic field considered in Section 3.

\subsection{Comparable toroidal and poloidal magnetic fields at the characteristic scale of MS mode. }
In that case  the conventional asymptotical method of subsequent interactions may be applyied:
\begin{equation}
\bar{S}(r)=\sqrt{\frac{\bar{\beta}_z(r)}{\bar{\beta}_\theta(r)}}\sim\beta^0,\,\,\,\tilde
{\beta}_z(r)=\frac{\bar{\beta}_z(r)}{\beta}\sim\beta^0,\,\,\,
\label{A17}
\end{equation}
\begin{equation}
\hat{b}_\theta(\eta)=\hat{b}_{\theta,0}(\eta)+
\beta^{-1}\hat{b}_{\theta,1}(\eta)+\dots,\,\,\,
 \hat{v}_z(\eta)=\hat{v}_{z,0}(\eta)+
\beta^{-1}\hat{v}_{z,1}(\eta)+\dots,\,\,\,
\lambda=\lambda_0+\beta^{-1 }\lambda_1+\dots.\,\,\,
\label{A18}
\end{equation}
Substituting (A17)-(A18) in (A1) -  (A4), and applying the principle of the least possible degeneration of the problem yield that the equation for $\hat{v}_{z,0}$ and $\hat{b}_{\theta,0}(\eta)$ are reduced to the conventional relation (\ref{A6})
for MS mode with the eigenvalues
$\lambda_0=\sqrt{m+1}$, ($m=1,2,\dots$, see Eq. (\ref{69})) 
\begin{equation}
\frac{d^2 \hat{v}_{z,0}}{d \eta^2 }-\eta\frac{d \hat{v}_{z,0}}{d \eta }+(\lambda_0^2-1)\hat{v}_{z,0}=0,
\label{A19}
\end{equation}
and to the relation for the MS-driven AC mode
\begin{equation}
\hat{b}_{\theta,0}=-\frac{3}{2\lambda_0^2}\bar{S}(r)\frac{d \hat{v}_{z,0}}{d \eta }.
\label{A20}
\end{equation}
Thus, on the characteristic scale of MS's mode, the stable MS mode decouples from the AC mode at large plasma beta, and the influence of the toroidal magnetic field is reduced to excitation of the stable MS-driven AC mode.

Finally note that above qualitative asymptotic analysis of the problem in the limit of large plasma beta demonstrates both spectral and transient stability of the disc embedded in the mixed equilibrium toroidal and poloidal magnetic fields of comparable values.



\bsp
\label{lastpage}

\begin{thebibliography}{99}
\bibitem[\protect\citeauthoryear{Balbus \& Hawley}{1991}]{Balbus and Hawley 1991}
Balbus  S. A.,  \& Hawley J. F., 1991,
ApJ,  {\bf 376},  214

\bibitem[\protect\citeauthoryear{Balbus \& Hawley}{1991}]{Balbus and Hawley 1992}
Balbus  S. A.,  \& Hawley J. F., 1992,
ApJ,  {\bf 400},  610

\bibitem[\protect\citeauthoryear{Begelman \& Pringle}{2007}]{Begelman and Pringle 2007}
Begelman  M.C. \& Pringle  J.E., 2007,
 MNRAS, {\bf 375}, 1070

\bibitem[\protect\citeauthoryear{Bodo et al.}{2008}]{Bodo et al. 2008}
Bodo  G., Mignone A., Cattaneo F., Rossi P., and Ferrari A.,
A\&A,  {\bf  487}, 1

\bibitem[\protect\citeauthoryear{Brandenburg et al.}{1995}]{Brandenburg et al. 1995}
Brandenburg A., Nordland A., Stein R. F., Torkelson U., 1995,
ApJ,  {\bf 446},  741


\bibitem[\protect\citeauthoryear{Chandrasechar}{1960}]{Chandrasechar 1960}
Chandrasechar S., 1960, Proc. Natl. Acad. Sci. A,  {\bf 46}, 46, 223

\bibitem[\protect\citeauthoryear{Coppi \& Keyes}{2003}]{Coppi and Keyes 2003}
Coppi  B., \& Keyes  E.A., 2003,
ApJ, {\bf 595}, 1000

\bibitem[\protect\citeauthoryear{Foglizzo \&  Tagger }{1995}]{Foglizzo and  Tagger  1995}
Foglizzo T., and  Tagger M., 1995,
A\&A,  {\bf  301}, 293


\bibitem[\protect\citeauthoryear{Frank  et al.}{2002}]{Frank  et al. 2002}
Frank  J., King  A., and Raine  D., 2002, \emph{Acreation Power in Astrophysics },
(Cambridge: University Press).


\bibitem[\protect\citeauthoryear{Fromang \&  Papaloizou }{2007}]{Fromang and Papaloizou  2007}
Fromang S.,  and Papaloizou J., 2007,
A\&A,  {\bf  476}, 1113


\bibitem[\protect\citeauthoryear{Fromang et al.}{2007}]{Fromang et al. 2007}
Fromang S., Papaloizou J., Lesur G., Heineman T.,  2007,
A\&A,  {\bf  476}, 1123


\bibitem[\protect\citeauthoryear{Gammie  \& Balbus }{1994}]{Gammie and Balbus 1994}
Gammie C. F., \&  Balbus  S. A., 1994,
 MNRAS, {\bf 270}, 138


\bibitem[\protect\citeauthoryear{Hawley  et al.}{1996}]{Hawley  et al. 1996}
Hawley  J. F., Gammie C. F., Balbus  S. A.,    1996,
ApJ, {\bf 464}, 690



\bibitem[\protect\citeauthoryear{Hawley \& Krolik }{2002}]{Hawley and Krolik  2002}
Hawley  J. F.,  \& Krolik  J. H., 2002,
ApJ, {\bf 566}, 164


\bibitem[\protect\citeauthoryear{Landau \& Lifshits}{1997}]{Landau and Lifshits 1997}
Landau  L. D. and Lifshits  E. M., 1977,
 \emph{Quantum mechanics, Non-Relativistic Theory },
(N.-Y.: Pergamon  Press).


\bibitem[\protect\citeauthoryear{Latter et al.}{2010}]{Latter et al. 2010}
Latter H. N., Fromang S., Gressel O., 2010
 MNRAS, {\bf 406}, 848

\bibitem[\protect\citeauthoryear{Lesur \& Longaretti}{2007}]{Lesur  and Longaretti 2007}
Lesur G., and Longaretti P. Y., 2007,
 MNRAS, {\bf 378}, 1471

\bibitem[\protect\citeauthoryear{Liverts \& Mond}{2009}]{Liverts and Mond 2009}
Liverts  E. and Mond  M., 2009,
 MNRAS, {\bf 392}, 287


\bibitem[\protect\citeauthoryear{Migdal}{1977}]{Migdal 1977}
Migdal  A. B., 1977,
 \emph{Qualitative Methods in Quantum Physics},
(Massachusetts: W. A. Bnejamin Inc.).


\bibitem[\protect\citeauthoryear{Nayfeh}{1973}]{Nayfeh 1973}
Nayfeh   A. H.,  1973,
 \emph{Perturbation Methods},
(New-York: John Willey \& Sons).

\bibitem[\protect\citeauthoryear{Papaloizou \&  Terquem}{1997}]{Papaloizou and Terquem 1997}
Papaloizou  J. C. B., \&  Terquem  C., 1997,
MNRAS, {\bf 287}, 771

\bibitem[\protect\citeauthoryear{Pessah \& Psaltis}{2005}]{Pessah and Psaltis 2005}
Pessah  M.E. and Psaltis  D., 2005,
ApJ,  {\bf 628},  829

\bibitem[\protect\citeauthoryear{Pessah  et al.}{2007}]{Pessah  et al.  2007}
Pessah M. E., Chan C. K., Psaltis D.,  2007,
 ApJ, {\bf 668}, L51

\bibitem[\protect\citeauthoryear{Proga }{2003}]{Proga  2003}
Proga  D., 2003,
ApJ, {\bf  585} 406


\bibitem[\protect\citeauthoryear{Rebusco et al.}{2009}]{Rebusco et al.  2009}
Rebusco   P., Umurhan  O. M., Kluzniak  W., and Regev  O., 2009,
Phys. Fluids,  {\bf 21}, 076601


\bibitem[\protect\citeauthoryear{Regev \&  Umurhan }{2008}]{Regev and Umurhan 2008}
 Regev  O., and Umurhan  O. M., 2008,
A\&A,  {\bf  481}, 21


\bibitem[\protect\citeauthoryear{Rudiger et al.}{2007}]{Rudiger et al.  2008}
Rudiger   G., Hollerbach  R., Schultz  M., and Detlev  E., 2007,
MNRAS, {\bf 377}, 1481


\bibitem[\protect\citeauthoryear{Schmidt \& Henningson}{2001}]{Schmidt and Henningson 2001}
Schmidt P. J., and Henningson D. S., 2001,
 \emph{Stability and transition in Shear Flows },
(Berlin: Springer).


\bibitem[\protect\citeauthoryear{Shtemler et al.}{2009}]{Shtemler et al. 2009}
Shtemler  Y. M.,  Mond  M., \& Rudiger  G., 2009,
 MNRAS, {\bf 394}, 1379


\bibitem[\protect\citeauthoryear{Shtemler et al.}{2010}]{Shtemler et al. 2010}
Shtemler  Y. M., Mond M., Rudiger G, Regev  O., \& Umurhan  O.M., 2010,
 MNRAS, {\bf 406}, 517


\bibitem[\protect\citeauthoryear{Spitzer}{1942}]{Spitzer 1942}
Spitzer L., 1942,
ApJ,  {\bf 95}, 329



\bibitem[\protect\citeauthoryear{Terquem  \& Papaloizou }{1996}]{Terquem  and Papaloizou 1996}
Terquem  C.,  \&  Papaloizou  J. C. B., 1996,
MNRAS, {\bf 279}, 767


\bibitem[\protect\citeauthoryear{Velichov}{1959}]{Velichov 1959}
Velichov E. P., 1959,
Zh.  Eksp. Teor. Fiz., {\bf  36}, 1398 \ \ [English translation, 1959, Sov. Phys. JETP, 36, 95]

\bibitem[\protect\citeauthoryear{Umurhan et al.}{2006}]{Umurhan et al. 2006}
Umurhan  O. M., Nemirovsky  A., Regev  O.,  \&  Shaviv  G., 2006,
A\&A,  {\bf  446}, 1








\end{thebibliography}
\end{document}